\documentclass[useAMS,usenatbib]{mn2e} 
\usepackage{color}
\usepackage{epsfig}
\usepackage{amsbsy}  
\usepackage{amssymb}

\newcommand{\be}{\begin{equation}} 
\newcommand{\ee}{\end{equation}}
\newcommand{\bea}{\begin{eqnarray}} 
\newcommand{\eea}{\end{eqnarray}}

\newcommand{\COSMOMC}{{\sc cosmomc}}
\newcommand{\CAMB}{{\sc camb}}

\def\lt{<}

\def\G{{\rm\thinspace G}}
\def\Hz{{\rm\thinspace Hz}}
\def\cm{{\rm\thinspace cm}}
\def\erg{{\rm\thinspace erg}}
\def\s{{\rm\thinspace s}}
\def\keV{{\rm\thinspace keV}}
\def\km{{\rm\thinspace km}}
\def\Mpc{{\rm\thinspace Mpc}}
\def\arcmin{{\rm\thinspace arcmin}}

\def\K{{\rm\thinspace K}}
\def\micro{{\rm\thinspace $\mu$}}
\def\ergps{\hbox{$\erg\s^{-1}\,$}}
\def\ergpcmsqps{\hbox{$\erg\cm^{-2}\s^{-1}\,$}}
\def\kmps{\hbox{$\km\s^{-1}\,$}}
\def\kmpspMpc{\hbox{$\kmps\Mpc^{-1}$}}
\def\spose#1{\hbox to 0pt{#1\hss}}
\def\approxpropto{\mathrel{\spose{\lower 3pt\hbox{$\sim$}}
	\raise 2.0pt\hbox{$\propto$}}}
\newcommand{\lsim}{\lesssim}
\newcommand{\gsim}{\gtrsim} \def \lleq {\lower0.9ex\hbox{$\buildrel \lt \over \sim$} ~}

\title[Prospects for X-ray gas mass fraction studies] 
{The prospects for constraining dark energy with future X-ray cluster 
gas mass fraction measurements}
\author[D.~Rapetti, S.~W.~Allen \& A.~Mantz] {David~Rapetti${}^{1}$\thanks{Email: drapetti@slac.stanford.edu},
  Steven~W.~Allen${}^{1}$ and Adam~Mantz${}^{1}$\\
  ${}^1$ Kavli Institute for Particle Astrophysics and Cosmology at\\
  Stanford University, 382 Via Pueblo Mall, Stanford 94305-4060, CA, USA, and\\
  Stanford Linear Accelerator Center, 2575 Sand Hill Road, Menlo Park 94025, CA, USA.}

\begin{document}
\date{Accepted ???, Received ???; in original form \today}
\pagerange{\pageref{firstpage}--\pageref{lastpage}} \pubyear{2007}

\maketitle
\label{firstpage}

\begin{abstract}
  We examine the ability of a future X-ray observatory, with
  capabilities similar to those planned for the Constellation-X or
  X-ray Evolving Universe Spectroscopy (XEUS) missions, to constrain
  dark energy via measurements of the cluster X-ray gas mass fraction,
  $f_{\rm gas}$. We find that $f_{\rm gas}$ measurements for a sample
  of $\sim500$ hot ($kT\gsim5$keV), X-ray bright, dynamically relaxed
  clusters, to a precision of $\sim 5$ per cent, can be used to
  constrain dark energy with a Dark Energy Task Force (DETF; Albrecht
  et al. 2006) figure of merit of $15-40$, with the possibility of
  boosting these values by 40 per cent or more by optimizing the
  redshift distribution of target clusters. Such constraints are
  comparable to those predicted by the DETF for other leading, planned
  `Stage IV' dark energy experiments.  A future $f_{\rm gas}$
  experiment will be preceded by a large X-ray or SZ survey that will
  find hot, X-ray luminous clusters out to high redshifts.  Short
  `snapshot' observations with the new X-ray observatory should then
  be able to identify a sample of $\sim 500$ suitably relaxed
  systems. The redshift, temperature and X-ray luminosity range of
  interest has already been partially probed by existing X-ray cluster
  surveys which allow reasonable estimates of the fraction of clusters
  that will be suitably relaxed for $f_{\rm gas}$ work to be made;
  these surveys also show that X-ray flux contamination from point
  sources is likely to be small for the majority of the targets of
  interest. Our analysis uses a Markov Chain Monte Carlo method which
  fully captures the relevant degeneracies between parameters and
  facilitates the incorporation of priors and systematic uncertainties
  in the analysis. We explore the effects of such uncertainties for
  scenarios ranging from optimistic to pessimistic. We conclude that
  the $f_{\rm gas}$ experiment offers a competitive and complementary
  approach to the best other large, planned dark energy experiments.
  In particular, the $f_{\rm gas}$ experiment will provide tight
  constraints on the mean matter and dark energy densities, with a
  peak sensitivity for dark energy work at redshifts midway between
  those of supernovae and baryon acoustic oscillation/weak
  lensing/cluster number counts experiments. In combination, these
  experiments should enable a precise measurement of the evolution of
  dark energy.
\end{abstract}


\section{Introduction}
\label{introduction}

In the early 1990s, measurements of the baryonic mass fraction in
X-ray luminous galaxy clusters provided compelling evidence that we
live in a low density Universe. Under the assumption that large
clusters provide approximately fair samples of the matter content of
the Universe, X-ray observations require that the mean matter density,
$\Omega_{\rm m}$, is significantly less than the critical value, with
a best-fit value $\Omega_{\rm m}\sim 0.2-0.3$
\citep[e.g.][]{White:91,Fabian:91, Briel:92, White:93, David:95,
  White:95, Evrard:97, Mohr:99, Ettori:99, Roussel:00, Grego:01,
  Allen:02, Allen:04, Allen:07, Ettori:03, Sanderson:03, Lin:03,
  LaRoque:06}. When combined with the expectation from inflation
models, later confirmed by Cosmic Microwave Background (CMB) studies
\citep[][and references therein]{Bennett:03, Spergel:03}, that the
Universe should be close to spatially flat, X-ray results on the
cluster baryon mass fraction quickly lead to the suggestion that the
mass-energy density of the Universe may be dominated by a cosmological
constant \citep[e.g.][]{White:93}.

The first direct evidence for late-time cosmic acceleration, as would
be produced by a sizeable cosmological constant, was provided in the
late 1990s by \cite{Riess:98} and \cite{Perlmutter:98} based on
measurements of the light curves of type Ia supernovae (SNIa). Since
then, larger SNIa data sets \citep{knop:03, Riess:04, Astier:06,
  Riess:07, WoodVasey:07, Davis:07} and an increasingly wide array of
other, complementary experiments have confirmed and improved upon this
striking measurement. The combination of CMB data from the Wilkinson
Microwave Anisotropy Probe (WMAP) \citep{Spergel:03, Spergel:06,
  Dunkley:08} with large scale structure (LSS) data from the Sloan
Digital Sky Survey (SDSS) \citep{Eisenstein:05, Percival:07} and/or
2dF Galaxy Redshift Survey (2dFGRS) \citep{Cole:05} provides powerful
evidence for dark energy. The cross-correlation of CMB and LSS
fluctuations reveals the effects of dark energy on the Integrated
Sachs-Wolfe effect \citep{Scranton:03, Fosalba:03, Rassat:07}.
Measurements of the amplitude and evolution of matter fluctuations
using X-ray galaxy clusters \citep{Borgani:01,Reiprich:02, Allen:03,
  Schuecker:03, Voevodkin:04, Henry:04, Mantz:07}, optically-selected
clusters \citep{Gladders:06, Rozo:07}, Lyman-$\alpha$ forest data
\citep{Viel:04, Seljak:04}, and weak lensing \citep{vanWaerbeke:05,
  Jarvis:05, Hoekstra:06, Benjamin:07}, also provide important,
powerful confirmation of the new, standard cosmological paradigm: a
universe in which the main mass and energy components are dark matter
and dark energy, and where dark energy drives the current
acceleration. The standard model for dark energy remains the
cosmological constant, which is mathematically equivalent to vacuum
energy. In principle, however, cosmic acceleration could be driven by
either dark energy or a modification to the laws of gravity on
cosmological scales \citep[see][for an extensive review]{Copeland:06}.

Building on the early X-ray work, \cite{Allen:04, Rapetti:05}; and
\cite{Allen:07} showed that measurements of the evolution of the X-ray
gas mass fraction, $f_{\rm gas}$, in the largest, dynamically relaxed
galaxy clusters provides a further powerful, complementary approach
for studying dark energy. As with SNIa data, $f_{\rm gas}(z)$
measurements probe the redshift-distance relation; whereas the peak
SNIa luminosity varies as the square of the distance, $f_{\rm gas}$
measurements vary as distance, $d^{1.5}$. \citep[The distance
dependance derives from the way in which $f_{\rm gas}$ values are
determined from the observed X-ray temperature and surface brightness
data;][]{Allen:07} In combination with the tight constraint on
$\Omega_{\rm m}$ provided by the normalization of the $f_{\rm gas}(z)$
curve, under the assumption of fair matter samples, the $f_{\rm
  gas}(z)$ data contain sufficient information to break the degeneracy
between $\Omega_{\rm m}$ and the dark energy equation of state, $w$,
in the distance equations. The additional combination of $f_{\rm gas}$
and CMB data breaks other important degeneracies between parameters in
cosmological analyses \citep{Rapetti:05, Allen:07}.

\cite{Allen:07} show that the current constraints on dark energy from
the $f_{\rm gas}$ experiment are of comparable precision to other
leading techniques, and are robust under the inclusion of conservative
systematic allowances, e.g. relaxing the requirement for exact
hydrostatic equilibrium and allowing for moderate redshift evolution
in the cluster baryon fraction. These authors also show that
intrinsic, systematic scatter remains undetected in the current
$f_{\rm gas}$ data, despite a weighted mean statistical scatter in the
individual distance measurements of only $\sim 5$ per
cent; in contrast, SNIa studies \citep{Riess:07, Jha:07, WoodVasey:07}
have established the presence of systematic scatter at the $\sim 7$
per cent in distance measurements from the best current SNIa data.

The key to determining the nature of dark energy is to 
obtain precise measurements of its evolution 
with redshift, $z$, or scale factor, $a=1/(1+z)$.  The Dark
Energy Task Force report \cite[][hereafter DETF]{Albrecht:06}
presented estimates of the constraints on dark energy parameters that
should be achievable with a number of future proposed or planned dark
energy experiments. In particular, the report forecasted the ability
of these experiments, in combination with CMB data from the Planck
satellite, to constrain a dark energy model of the form $w(a)=w_{\rm
0}+w_{\rm a}(1-a)$, and defined a figure of merit (hereafter FoM) to
allow for easy comparison of the constraints.  In this paper, we use
the same dark energy parameterization and FoM to quantify the
constraining power of future $f_{\rm gas}$ experiments, to be carried
out with e.g. the Constellation-X or X-ray Evolving Universe
Spectroscopy (XEUS) missions, in combination with CMB data. We show
that the $f_{\rm gas}$ experiment is likely to provide comparable
constraining power to the best other, contemporary space and
ground-based experiments described by the DETF. When combined, 
future CMB, SNIa, baryon acoustic oscillation (BAO), weak
lensing, cluster number count and $f_{\rm gas}$ experiments should
provide precise, accurate constraints on $w(z)$ and allow significant
progress in understanding the origin of cosmic acceleration.

The structure of this paper is as follows: in Section~\ref{sec:de} we
define the dark energy model and the FoM. In Section~\ref{sec:simdata}
we describe the simulated $f_{\rm gas}$ and CMB data sets.  For the
$f_{\rm gas}$ data, we assume instrument characteristics appropriate
for the baseline Constellation-X mission. The CMB data set
approximates that expected from two years of Planck data. We also
simulate a data set representative of that produced by follow-up
observations of the Sunyaev-Zel'dovich effect in the clusters targeted
for the $f_{\rm gas}$ work.  Section~\ref{analysis} describes the
Markov Chain Monte Carlo (MCMC) pipeline and details of the analysis
method. Our main results are presented in
Section~\ref{constraints}. Section~\ref{conclusions} summarizes our
conclusions.

\section{The dark energy model and FoM}
\label{sec:de}

We characterize the evolution of dark energy by its energy density in
units of the critical density, $\Omega_{\rm de}$, and its equation of
state, $w$.  Following the DETF, we parameterize the evolution of the
dark energy equation of state as $w(a)=w_{\rm 0}+w_{\rm a}(1-a)$
\citep{Chevallier:01, Linder:03} for which a cosmological constant has
$w(a)=-1$. In this model, the dimensionless Hubble parameter as a
function of scale factor has the form

\begin{equation}
  E(a)=\frac{H(a)}{H_{\rm 0}}=\sqrt{\Omega_{\rm m}a^{\rm -3}+\Omega_{\rm de}f(a)+\Omega_{\rm k}a^{\rm -2}}\,,
\label{eq:eoz}
\end{equation}

\noindent where

\begin{equation}
  f(a)=a^{-3(1+w_{\rm 0}+w_{\rm a})}e^{-3w_{\rm a}(1-a)}\,.
\label{eq:eoz2}
\end{equation}

\noindent $H_{\rm 0}$ is the present-day value of the Hubble parameter
and $\Omega_{\rm m}$ and $\Omega_{\rm k}$ are the mean matter density
and curvature density in units of the critical density, respectively.

Using this parameterization, the DETF define a FoM that is used to
compare the constraining power of different dark energy experiments.
Nominally, the FoM scales with the inverse of the area enclosed by the
95 per cent confidence contour in the $w_{\rm 0}-w_{\rm a}$ plane.
However, the DETF showed that since there is little correlation in the
$w_{\rm p}-w_{\rm a}$ plane, the area is also proportional to the
product of the standard deviations $\sigma(w_{\rm p}) \times
\sigma(w_{\rm a})$, where $w_{\rm p}=w(a_{\rm p})$ is the pivot value
of $w(a)$, i.e., the value of $w(a)$ at which its uncertainty is
minimized \citep{Linder:06b}. (Note that the standard error
$\sigma(w_{\rm p})$ approximately corresponds to the 68.3 per cent
uncertainty in $w$ that would be obtained for a constant$-w$ dark
energy model). This leads to the definition

\begin{equation}
{\rm FoM}=[\sigma(w_{\rm p})\times\sigma(w_{\rm a})]^{\rm -1}.
\label{eq:fom}
\end{equation}

\noindent For the DETF Fisher matrix analysis, the $1\sigma$
confidence region in the $w_{\rm p}-w_{\rm a}$ plane forms an ellipse
for which the semi-axes are the standard deviations of $w_{\rm p}$ and
$w_{\rm a}$. For the more detailed MCMC analysis used here, however,
we obtain slightly asymmetric probability distributions for these
parameters in some cases, although to either side of the peak
probability the distributions can be modelled as approximately
Gaussian.  Therefore, in calculating the FoM, we model the 1$\sigma$
confidence contour in the $w_{\rm p}-w_{\rm a}$ plane with a
geometrical shape formed by four quarters of four different ellipses
for which the semi-axes are the standard deviations of the Gaussians
to either side of the peak, namely $\sigma_{\rm up}(w_{\rm p})$,
$\sigma_{\rm down}(w_{\rm p})$, $\sigma_{\rm up}(w_{\rm a})$, and
$\sigma_{\rm down}(w_{\rm a})$. The area of such contour is equivalent
to the area of an ellipse with semi-axes $\hat{\sigma}(w_{\rm
  p})=[\sigma_{\rm up}(w_{\rm p})+\sigma_{\rm down}(w_{\rm p})]/2$ and
$\hat{\sigma}(w_{\rm a})=[\sigma_{\rm up}(w_{\rm a})+\sigma_{\rm
  down}(w_{\rm a})]/2$.  Thus, we calculate our FoM\footnote{To
  confirm the validity of our definition of the FoM we have explicitly
  measured the area contained by the filled contours in the right
  panel of Figure~\ref{fig:ws}.  Dividing this area by both the
  geometric factor $\pi$, which accounts for the conversion between
  the area of an ellipse and a quarter of its circumscript rectangle,
  and the factor $2.3$, which accounts for the change in the
  confidence levels from two to one degrees of freedom, we
  successfully match the measured area to the value obtained by the
  product $\hat{\sigma}(w_{\rm p})\times\hat{\sigma}(w_{\rm a})$.} as
the inverse of the product of the semi-axes $[\hat{\sigma}(w_{\rm
  p})\times\hat{\sigma}(w_{\rm a})]^{-1}$ which allows a direct
comparison with the results reported by the DETF.

\section{Simulated X-ray data}
\label{sec:simdata}

\subsection{A strategy for future $f_{\rm gas}$ work}
\label{fgasdata}

\begin{table}
\begin{center}
\caption{Baseline X-ray observatory characteristics.}
\label{tab:spec}
\begin{tabular}{ l l }
\hline

Band pass & 0.3-10 \keV \\  \noalign{\vskip 3pt}
Spectral resolution    & $E/\Delta E \sim 2400$ (@$6$ \keV) \\  \noalign{\vskip 3pt}
Effective area         & $15,000$ \cm$^2$ (@$1.25$ \keV)\\  \noalign{\vskip 3pt}
PSF                    & $\leq 15$ arcsec (half power diameter) \\  \noalign{\vskip 3pt}
Field of View          & $\geq 5\times 5$ arcmin$^2$ \\  \noalign{\vskip 3pt}
 
\hline
\end{tabular}
\end{center}
\end{table} 

We assume that a future $f_{\rm gas}$ experiment will be carried out
by an X-ray observatory with capabilities comparable to those of
Constellation-X, as summarized in Table 1. The major improvements of
such a mission with respect to current X-ray observatories are in
collecting area, which is a factor $\sim 100$ larger than that
provided by the Chandra X-ray Observatory, and spectral
resolution.\footnote{For details on planned X-ray observatories see
  http://constellation.gsfc.nasa.gov/ and
  http://www.rssd.esa.int/index.php?project=XEUS.}  We assume that the
$f_{\rm gas}$ experiment will be preceded by, and will build upon,
forthcoming X-ray and/or SZ cluster surveys\footnote{Forthcoming X-ray
  survey missions include Spectrum-RG/eROSITA; see
  http://www.mpe-garching.mpg.de/projects.html\#erosita and
  http://www.mpe-garching.mpg.de/erosita/MDD-6.pdf. Several large-area
  SZ surveys are already underway, including the South Pole Telescope
  (SPT) \citep[e.g.][see http://spt.uchicago.edu/]{SPT:04}, and the
  Atacama Cosmology Telescope (ACT) \citep[e.g.][see
  http://wwwphy.princeton.edu/act/]{Sehgal:07}.}  that will scan a
significant fraction of the sky and find a large number of hot, X-ray
luminous, high$-z$ clusters.  These surveys will provide the initial
target lists for the $f_{\rm gas}$ experiment as well as allowing an
array of complementary cosmological tests based on the power spectrum
and mass function of galaxy clusters \citep[e.g.][]{Albrecht:06}.

From initial surveys of tens of thousands of clusters, the $\sim 4000$
most X-ray luminous (or highest integrated SZ flux) clusters will be
identified. The new X-ray observatory will then be used to take short
snapshot exposures ($\sim 1$ks) of these clusters, to identify the
most apparently dynamically relaxed systems that are most suitable for
$f_{\rm gas}$ work \citep{Allen:07}. The selection of relaxed clusters
is likely to be based primarily on X-ray morphology, but will also
utilize the high spectral resolution capabilities to measure bulk gas
motions.\footnote{The snapshot observations will also be of great
  benefit for a range of ancillary cluster science.} The most relaxed
clusters will be re-observed with deeper exposures to measure the gas
mass fraction to the required level of precision.

Current studies of the Massive Cluster Survey (MACS)
\citep{Ebeling:01,Ebeling:07} show that at redshifts $z\lsim0.5$
approximately $1/4$ clusters are sufficiently relaxed for $f_{\rm
  gas}$ work \citep{Allen:07}. We (conservatively) calculate predicted
cosmological constraints for two separate $f_{\rm gas}$ data sets,
containing either $\sim 500$ or $250$ relaxed clusters.  That is, we
assume that only approximately $1/8$ or $1/16$ of the 4000 hottest,
most X-ray luminous clusters detected in a future survey will be
suitable for use in the $f_{\rm gas}$ experiment.

For the $500-$cluster sample, we assume an average exposure time per
cluster of $\sim 20$ks. For the $250-$cluster sample, the typical
exposure is $\sim 40$ks. In both cases, the total time required to
complete the $f_{\rm gas}$ observations will be $\lsim 15$Ms. For the
assumed instrument characteristics, we expect statistical
uncertainties in the $f_{\rm gas}$ measurements resulting from $20$ks
exposures of $\sim 5$ per cent, which corresponds to $\sim 3.3$ per
cent in distance. For typical exposures of $40$ks, we expect to
measure $f_{\rm gas}$ to $\sim 3.5$ per cent or distance to $\sim 2.3$
per cent.  In Section~\ref{constraints} we show that the constraints
on dark energy from both the 500 or 250-cluster sample are
comparable. We adopt the $500-$cluster sample with $5$ per cent
$f_{\rm gas}$ measurement uncertainties as our default data set.

\subsection{The simulated $f_{\rm gas}$ data set}

\subsubsection{The luminosity function of clusters}
\label{sec:lf}

To simulate the $f_{\rm gas}$ data set, we first need to predict the
redshift distribution of clusters.  We assume an X-ray flux-limited
cluster survey similar to that expected to be produced by the
Spectrum-RG/eROSITA mission, with a flux limit of $F_{\rm
  lim}=3.3\times10^{-14}$\ergpcmsqps in the $0.1-2.4\keV$ band and a
uniform sky coverage of $f_{\rm sky}=0.5$. We calculate the number of
clusters expected to be observed, $N_{\rm i}$, in each redshift bin,
$z_{\rm i}$, as \citep{Mantz:07}

\begin{equation}
  N_{\rm i}(z_{\rm i})=\int^{z_{\rm i}}_{z_{\rm i-1}}\frac{dV}{dz}\,dz\int^{\infty}_{0}\frac{dn(M,z)}{dM} Q \,dM\,,
\label{eq:nofz}
\end{equation} 

\noindent where 

\begin{equation}
Q=\int^{\infty}_{0}dL'\,\int^{\infty}_{L_{\rm lim}(z)}dL\, P(L'|M)\, p(L|L').
\label{eq:nofz2}
\end{equation} 

\noindent Here, $V$ is the comoving volume, $n(M,z)$ is the comoving
number density of halos with a mass less than $M$ at redshift $z$,
$L'$ is the intrinsic luminosity of a galaxy cluster associated with a
halo of mass $M$, and $L$ is its luminosity inferred from
observations.  $P(L'|M)$ is the probability for a cluster of mass $M$
to have an intrinsic luminosity $L'$; $p(L|L')$ is the probability for
a cluster with intrinsic luminosity $L'$ to be observed with
luminosity $L$; and $L_{\rm lim}(z)$ is the luminosity limit
function. We calculate the comoving volume element per redshift
interval as \citep{Hogg:99}
 
\begin{equation}
  \frac{dV}{dz}= 4\pi\,f_{\rm sky}\,\frac{c}{H_{\rm 0}}\,\frac{(1+z)^2\,d_{\rm A}(z)^2}{E(z)}\,,
\label{eq:dvdz}
\end{equation}

\noindent where $c$ is the speed of light, and $d_{\rm A}$ the angular
diameter distance. Using N-body simulations, \cite{Jenkins:01} obtained the following
fitting formula for the mass function of dark matter halos:

\begin{equation}
  {{\rm d}n(M, z)\over{\rm d}\ln\sigma^{-1}} = \frac{\bar{\rho}}{M}\,A \exp\left[-|\ln\sigma^{-1}+ B\,|^{\epsilon}\right] \,,
\label{massj}
\end{equation}

\noindent where $\bar{\rho}$ is the comoving mean matter density of
the Universe and $A$, $B$ and $\epsilon$ are fitted parameters. Here
$\sigma^2(M,z)$ is the variance of the linearly evolved density field,
smoothed by a spherical top-hat filter, $W(k;M)$. In Fourier-space
representation,

\begin{equation}
  \sigma^2(M,z) = {D^2(z)\over2\pi^2}\int_{\rm 0}^\infty k^2P(k)W^2(k;M){\rm d}k,
\label{variance}
\end{equation} 

\noindent where $k$ is the wave number, $P(k)$ is the power spectrum
of the linear density field extrapolated to redshift zero and $D(z)$
is the growth factor of linear perturbations normalized to be $1$ when
$z=0$. We calculate the power spectrum using the {\sc CAMB} code
\citep{Lewis:00}\footnote{http://camb.info/}. For halo finding
algorithms tied to the mean mass density, \cite{Jenkins:01} showed
that the values of $A$, $B$ and $\epsilon$ are almost invariant under
both a broad range of cosmologies and redshift. However, these authors
also showed that these parameters depend on the cluster finding
algorithm. Here, we use $A=0.316$, $B=0.67$, $\epsilon=3.82$
\citep{Jenkins:01}, which are appropriate for the spherical
overdensity algorithm SO($\kappa=324$) \citep{Davis:85, Lacey:94},
where $\kappa$ is the mean overdensity of the halo with respect to the
mean matter density of the Universe.

In equation (\ref{eq:nofz2}) we have a log-normal probability
distribution \citep{Mantz:07} 

\begin{equation}
P(L'|M)= \frac{e^{[\log_{10}L'-\log_{10}\hat{L'}(M)]^2/2\sigma^2}}{L'\ln(10)\sqrt{2\pi}\sigma}\,,
\label{eq:lognormal}
\end{equation}

\noindent where $\hat{L'}(M)$ is the best fit luminosity for a given
mass $M$, and $\sigma$ is its scatter, determined from the
mass-luminosity data set of \cite{Reiprich:02} using the relation

\begin{equation}
  \log_{10}\left[\frac{M\,E(z)}{h_{72}^{-1}M_{\sun}}\right]=\mathcal{A}+\alpha\log_{10}\left[\frac{L_{\rm X}(0.1-2.4\keV)}{10^{44}h_{72}^{-2}\ergps E(z)}\right]\,,
\label{eq:MLrel}
\end{equation}

\noindent for which $\alpha=0.67$ and
$\mathcal{A}=\log_{10}\left[M_{\rm
    0}/(h_{72}^{-1}M_{\sun})\right]=14.49$, and $\sigma=0.12$, as
obtained by \cite{Mantz:07}. In equation (\ref{eq:nofz2}) we also have
a Gaussian probability distribution

\begin{equation}
p(L|L')= \frac{e^{[L-L']^2/2\sigma_{\rm l}^2}}{\sqrt{2\pi}\sigma_{\rm l}}\,,
\label{eq:normal}
\end{equation}

\noindent with standard deviation $\sigma_{\rm l}=(\sigma_{\rm n_{\rm
    ph}}/n_{\rm ph})L$. Here $n_{\rm ph}$ is the number of photons
detected from a cluster in the survey and $\sigma_{\rm n_{\rm
    ph}}=\sqrt{n_{\rm ph}}$ is the associated Poisson error. We assume
that at the flux limit of the survey, $F_{\rm lim}$, or equivalently
at the luminosity limit $L_{\rm l}=L(F_{\rm lim},z)$, the number of
photons is $n_{\rm ph,lim}\sim 20$. Using this, we have $\sigma_{\rm
  l}=(\sqrt{L_{\rm l}/n_{\rm ph,lim}})\sqrt{L}$.

\subsubsection{Temperature selection}
\label{sec:selec}

In order to minimize systematic scatter in the $f_{\rm gas}$
experiment, \cite{Allen:07} restrict their analysis to dynamically
relaxed clusters with mean gas mass-weighted temperatures measured
within $r_{2500}$,\footnote{$r_{2500}$ is the radius within which the
  mean density is 2500 times the critical density of the Universe at
  the redshift of the cluster.}~$kT_{\rm 2500}>5$\keV.  We impose the
same temperature cut in this analysis, calculating the luminosity
limit $L_{\rm i}$ that corresponds to this temperature limit from the
relation \citep{Bryan:98}
 
\begin{equation}
\log_{10}\left[\frac{L_{\rm X}(0.1-2.4\keV)}{10^{44}h_{72}^{-2}\ergps E(z)}\right]=A+B\log_{10}\left(\frac{kT_e}{\keV}\right)\,,
\label{eq:LTrel}
\end{equation}

\noindent where $T_e$ is the emission weighted X-ray
temperature~\footnote{$T_e$ scales with $T_{2500}$ as $kT_{\rm e} \sim
  kT_{2500}/\eta$ with $\eta\sim1.1-1.2$, based on MACS clusters
  spanning the redshift range $0.3<z<0.7$. Beyond redshift $0.7$ the
  value of $\eta$ slowly decreases towards $\sim 1$. To be
  conservative, however, we ignore the difference between $T_e$ and
  $T_{2500}$, i.e. we assume $\eta=1$, at all redshifts.}. Fitting the
above relation (\ref{eq:LTrel}) to the X-ray luminosity and
temperature data of \cite{Reiprich:02} using the linear regression
BCES(Y$|$X) algorithm of \cite{Akritas:96}, we obtain $A=-1.46 \pm
0.09$ and $B=2.50 \pm 0.13$.\footnote{We exclude objects from the
  \cite{Reiprich:02} sample for which the temperature was estimated
  from the luminosity-temperature relation of \cite{Markevitch:98}
  rather than directly measured. This leaves 88 data points in total.}

The limiting luminosity in equation (\ref{eq:nofz2}) is then

\begin{equation}
  L_{\rm lim}(z)={\rm min}[L_{\rm i},\,4\pi F_{\rm lim} d_{\rm L}(z)^2]\,, 
\label{eq:llim}
\end{equation}

\noindent with the appropriate $K-$correction applied in 
calculating the $F_{\rm lim}$ values.

\subsubsection{The redshift distribution of $f_{\rm gas}$ clusters}
\label{sec:fgas_distr}

\begin{figure}
\includegraphics[width=3.2in]{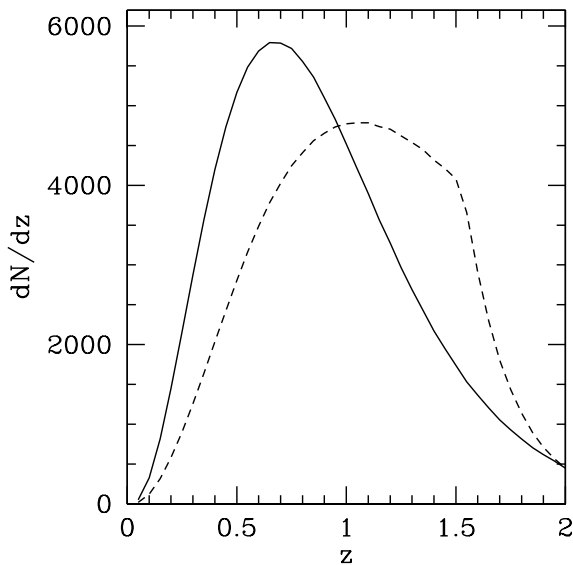}
\caption{The redshift distribution (solid curve) of clusters above the
  Spectrum-RG/eROSITA X-ray flux limit with temperatures
  $kT_{2500}>5$keV. A sky coverage of $50$ per cent is assumed. This
  redshift distribution has been used to generate the mock $f_{\rm
    gas}$ data set. Assuming that $\sim 1/8$ of such clusters will be
  sufficiently relaxed for $f_{\rm gas}$ work, we obtain a final
  sample of $\sim 500$ clusters.  For comparison purposes, we also
  show (dashed curve) the redshift distribution for the case of a
  fixed luminosity limit $L_{\rm X}(0.1-2.4$\keV$)>3.35\times
  10^{44}h_{\rm 70}^{-2}$\ergps (no temperature cut) which gives a
  similar total number of clusters. The latter distribution has more
  high$-z$ clusters.}
\label{fig:distr}
\end{figure}

Table~\ref{tab:fidu} summarizes the parameters describing our fiducial
cosmology. For this cosmology, we have calculated the redshift
distribution of galaxy clusters over the range $0<z<2$.  Our fiducial
cosmology approximately matches that used by the DETF, but includes
updated values for $n_{\rm s}$ and $\tau$ to better match the WMAP
three-year and five-year results \citep{Spergel:06, Dunkley:08}. We
also adopt a lower value for $\sigma_{8}=0.8$, consistent with both
the WMAP three-year and five-year results and the results of
\cite{Mantz:07} from measurements of the X-ray luminosity function of
galaxy clusters within $z<0.7$.

Figure~\ref{fig:distr} shows the redshift distribution (solid line)
for clusters detected above the Spectrum-RG/eROSITA X-ray flux limit
with mass-weighted temperatures $kT_{2500}>5$keV. A sky coverage,
$f_{\rm sky}=0.5$ is assumed. Approximately $5000$ clusters meet these
criteria from which, following our observing strategy, $4000$ will be
observed by short snapshots. Assuming that $\sim 1/8$ of these
clusters will also meet the relaxation criteria based on X-ray
morphology \citep{Allen:07, Million:08}, a sample of $\sim 500$ hot,
X-ray luminous, dynamically relaxed clusters can be defined. Taking
snapshot observations of the available $\sim 5000$ clusters instead of
$4000$, and assuming that $\sim 1/8$ of these clusters are relaxed, we
will obtain a sample of $\sim 625$ $f_{\rm gas}$ targets. This allows
us to either use a larger sample of clusters, assume an even more
conservative ratio of relaxed clusters, or select a different redshift
distribution for the $f_{\rm gas}$ sample of $\sim 500$ clusters. In
Section~\ref{redshift_distr}, we discuss the latter case.

For comparison purposes, Figure~\ref{fig:distr} also shows (dashed
curve) the redshift distribution for the case of a luminosity limit of
$L_{\rm i}> 3.35\times 10^{44}h_{\rm 70}^{-2}$\ergps in the
$0.1-2.4$\keV~band (dashed line; no temperature cut is imposed). The
effect of the X-ray flux limit on the distribution is evident towards
the highest redshifts ($z\sim1.5$) in this case.

It is clear from that figure that the temperature and luminosity cuts
lead to different redshift distributions~\footnote{The redshift
  distributions shown in Figure~\ref{fig:distr} are sensitive to the
  mass-observable relation obtained by \cite{Mantz:07} using current
  data. See that work for details.}. In the case of the temperature
cut (solid line), the redshift distribution peaks around $z\sim0.65$
and relatively few clusters are found at $z>1.5$. For the case of the
luminosity cut (dashed line), the distribution peaks around $z\sim 1$,
and has many more clusters in the redshift range $1<z<2$. It is
important to note, however, that a redshift distribution weighted
towards higher redshifts does not necessarily imply tighter
constraints on dark energy. For the DETF FoM criterion, constraints
around the pivot redshift are important; for the $f_{\rm gas}$
experiment $z_{\rm p}\sim 0.25$ (see Figure~\ref{fig:evol}). In
Section~\ref{redshift_distr} we further discuss the effect that using
different redshift distributions has on the dark energy constraints.

We generate mock $f_{\rm gas}$ measurements for 500 clusters with the
redshift distribution appropriate for the case of the temperature cut
[solid curve, Figure~\ref{fig:distr}; in accordance with the selection
criteria used for current $f_{\rm gas}$ work \citep{Allen:07}].  For
each cluster, we assign a statistical error in the $f_{\rm gas}$
measurements of $\sim 5$ per cent.  We have also generated a set of
mock measurements for the case of $250$ clusters observed with $f_{\rm
  gas}$ measurements accurate to $3.5$ per cent. This latter data set
is used to study the impact on the dark energy constraints in the case
that the fraction of suitably relaxed clusters is less than $1/8$ at
high redshifts.

We stress that the predicted redshift distribution, which peaks around
$z\sim 0.65$ in the case of the temperature cut, has already been
probed, at least partially, over the luminosity and temperature range
of interest, by the MACS survey \citep{Ebeling:01}; MACS covers the
redshift range $0.3<z<0.7$ to a flux limit of $F_{\rm
  lim}=10^{-12}$\ergpcmsqps~in the $0.1-2.4\keV$ band. For MACS,
approximately $1/4$ clusters are found to be sufficiently relaxed for
$f_{\rm gas}$ work \citep{Allen:07}. Therefore, our assumption that
$\sim 1/8$ clusters detected in a future X-ray survey and meeting the
X-ray flux and temperature criteria will be suitably relaxed, appears
reasonable. Moreover, as discussed in Section 5.1, for the case of the
250-cluster sample (i.e. assuming that only $\sim 1/16$ clusters are
relaxed) and using a similar total observing time to obtain individual
$f_{\rm gas}$ measurements to $\sim 3.5$ per cent accuracy, we obtain
very similar dark energy constraints (see Table~\ref{tab:models}).

A final important point regards contaminating point sources: for MACS
clusters, the fraction of the measured $0.1-2.4\keV$ X-ray flux
arising from contaminating point sources is small, typically of order
a per cent (Mantz et al. 2008; this is also the case for the hottest,
$kT_{\rm e}\gsim5$keV, relaxed clusters at lower redshifts.)
Therefore, we do not expect our target clusters, which have comparable
X-ray temperatures and luminosities, to be severely affected by
contaminating point sources, especially at $z\lesssim 1$. This
alleviates the instrumental requirements on the point spread function.
An instrument with capabilities similar to the baseline
characteristics listed in Table 1 should be capable of making
significant strides in dark energy work.

\begin{table}
\begin{center}
  \caption{Parameter values of our fiducial cosmology, which is a flat
    $\Lambda$CDM cosmology.}
\label{tab:fidu}
\begin{tabular}{ l l l }
\hline

$w_{0}=-1$          &  $\Omega_{m}=0.27$    &  $H_{0}=72$\kmpspMpc            \\  \noalign{\vskip 3pt}
$w_{a}=0$           &  $\Omega_{b}=0.046$   &  $n_{\rm s}=0.95$               \\  \noalign{\vskip 3pt}
$\Omega_{de}=0.73$  &  $\sigma_{8}=0.8$    &  $A_{\rm s}=2.3\times 10^{-9}$  \\  \noalign{\vskip 3pt}
$\Omega_{k}=0$      &  $b=0.82$            &  $\tau=0.09$                    \\  \noalign{\vskip 3pt}
 
\hline
\end{tabular}
\end{center}
\end{table}

\subsection{Follow-up SZ observations}

The thermal SZ effect is a modification to the CMB spectrum caused by
Compton scattering of CMB photons by hot electrons in the intracluster
medium. The SZ flux measured at radio or sub-mm wavelengths can be
expressed in terms of the Compton $y-$parameter. For a given
cosmology, the $y-$parameter can also be predicted from the same X-ray
data used to determine the $f_{\rm gas}$ measurements, being
proportional to the integral along the line-of-sight of the product of
electron density and temperature, $\int{n_{\rm e} T_{\rm e} dl}$.

We have examined the additional cosmological constraining power that
can be achieved with follow-up radio/sub-mm SZ observations of our
sample of 500 clusters, assuming direct SZ flux measurements accurate
to 2 or 5 per cent \cite[a level of accuracy that should be
straightforward for SZ detector technology available at the time of
the experiment; see][and references therein]{Muchovej:07}. The
statistical uncertainties in the predicted Compton $y-$parameters will
be comparable to those associated with the $f_{\rm gas}$ measurements:
$\sim 5$ per cent for the 500-cluster sample. We generate our
predicted $y-$parameter data set for the redshift distribution shown
in Fig. 1 (solid curve).

\subsection{Mock CMB data sets}
\label{planckdata}

We have used the {\sc CAMB} code \citep{Lewis:00} to generate auto and
cross temperature and polarization angular power spectra, $C^{\rm
  TT}_{\rm l}$, $C^{\rm TE}_{\rm l}$ and $C^{\rm EE}_{\rm l}$, for the
fiducial, flat $\Lambda$CDM cosmology described in
Table~\ref{tab:fidu}. We follow \cite{Lewis:05} and \cite{Lewis:06}
and assume that the temperature, $T$, and polarization $E-$fields are
Gaussian and isotropic. We also assume that the polarization $B-$field
is negligible.

Having $C^{\rm TT}_{\rm l}$, $C^{\rm TE}_{\rm l}$ and $C^{\rm EE}_{\rm
l}$, we add a simple, isotropic noise power spectrum \citep{Cooray:00,
Lewis:05, Lewis:06}

\begin{equation}
  N_{\rm l}(\nu',\nu)= \mathcal{N}_{\rm l}(\nu)\,\delta_{\rm \nu',\nu} \, e^{b(\nu)^2 l(l+1)/8\ln 2},
\label{npower}
\end{equation}

\noindent where $\mathcal{N}_{\rm l}(\nu)=[b(\nu) \sigma(\nu)]^2$,
$b(\nu)=\sqrt{8\ln 2} \sigma(\nu)$ is the beam full width at half
maximum (FWHM) measured in radians, and $\sigma(\nu)=(\Delta T/T)^2$ is
the root mean square noise per beam-sized pixel. Assuming uncorrelated
noise in the $E$ and $T$ fields, the covariance over realizations is
\citep{Lewis:06}

\begin{equation}
  C_{\rm l}=\left(\begin{array}{cc} 
      C^{\rm TT}_{\rm l}+N^{\rm TT}_{\rm l} & C^{\rm TE}_{\rm l} \\
      C^{\rm TE}_{\rm l} & C^{\rm EE}_{\rm l}+N^{\rm EE}_{\rm l} \\
    \end{array} \right)\,.
\label{cov}
\end{equation}

\noindent For the channel $\nu=143$\G\Hz~, $\mathcal{N}^{\rm TT}_{\rm
  l}(\nu)=\mathcal{N}^{\rm EE}_{\rm l}(\nu)/4=2\times
10^{-4}$\micro\K$^{2}$ and $b(\nu)=7.1$\arcmin~\citep{Lewis:06}. These
values correspond to $\sigma(\nu)^{\rm TT}=6.97$\micro\K~and
$\sigma(\nu)^{\rm EE}=9.68$\micro\K, which is roughly the sensitivity
expected for {\sc
  Planck}\footnote{http://www.rssd.esa.int/index.php?project=Planck}
after $\sim 2$ years (14 months) of a full sky survey
\citep{Planck:06}.

We consider two different scenarios relating to foreground CMB
polarization contamination. Firstly, we examine the idealized case
where such contamination can be neglected \citep{Bond:04, Lewis:05,
Lewis:06, Planck:06}. Secondly, we consider a more conservative
scenario where $\sim 20$ per cent of the sky is irretrievably contaminated by
Galactic emission, leaving $\sim 80$ per cent that can be modelled as
approximately foreground-free. For the second scenario,
\cite{Tegmark:00} forecast that {\sc Planck} will be able to determine
the optical depth to reionization to a precision of $\sigma(\tau) \sim
0.01$, as compared to $\sigma(\tau) \sim 0.005$ for the idealized,
foreground-free case \citep{Bond:04, Lewis:06, Planck:06}. To account
for the effects of polarization contamination, the DETF discarded
polarization information for multipoles $l<30$ and imposed a prior on
$\tau$ to obtain $\sigma(\tau)=0.01$. For our analysis in the case of
polarization contamination, we also artificially weaken the
constraints on $\tau$ to a precision of $\sigma(\tau)\sim 0.01$ by
enlarging by an order of magnitude the noise at low multipoles $l<30$
in the polarization data.

For both scenarios, we use only the data from multipoles $2\leq l\leq
2000$. For simplicity, we adopt the zero-contamination scenario as our
default CMB data set.

\section{Data analysis method}
\label{analysis}

\subsection{Markov Chain Monte Carlo (MCMC) code}

Given the dark energy model described in Section~\ref{sec:de} and the
simulated $f_{\rm gas}$ and CMB data sets described in
Section~\ref{sec:simdata}, we use the Metropolis Markov Chain Monte
Carlo (MCMC) algorithm implemented in the
\COSMOMC~\footnote{http://cosmologist.info/cosmomc/}~\citep{Lewis:02}
package to examine posterior parameter distributions. We use a
modified version of the \CAMB~\citep{Lewis:00} code to calculate CMB
power spectra; this accounts for the effects of dark energy
perturbations for evolving dark energy equations of state
\citep{Rapetti:05} (see Section~\ref{sec:depert} for details). Our
modified version of the \COSMOMC~code also incorporates the \ $f_{\rm
  gas}$ analysis method described by \cite{Allen:07} \citep[see
also][]{Rapetti:05, Rapetti:07}.

Our choice to forecast parameter constraints using a full MCMC
analysis has some advantages over the more widely used Fisher matrix
formalism \citep[see discussions in][]{Perotto:06,Lewis:06}.  Firstly,
the shape of the mean log likelihood [see equation (\ref{eq:logli})]
\citep{Lewis:06} in the MCMC analysis encapsulates all of the relevant
degeneracies between parameters, which is crucial for non-Gaussian
distributions. Secondly, the fact that our forecasts are made using
the same \COSMOMC~analysis code used to analyze current data
\citep{Allen:07} ensures consistency between present and future
constraints. Finally, the MCMC method allows us to easily and
efficiently introduce priors and allowances and thereby study the
effects of systematic uncertainties.

\subsection{X-ray gas mass fraction analysis}
\label{sec:cldata}

\subsubsection{The $f_{\rm gas}$ method}

The X-ray gas mass fraction, $f_{\rm gas}$, is defined as the ratio of
the X-ray emitting gas mass to the total mass of a cluster. This
quantity can be determined from the observed X-ray surface brightness
and the deprojected, spectrally-determined gas temperature profile,
under the assumptions of spherical symmetry and hydrostatic
equilibrium.  To ensure that these assumptions are as accurate as
possible, it is essential to limit the $f_{\rm gas}$ analysis to the
hottest, most X-ray luminous, dynamically relaxed clusters available
[Section~\ref{fgasdata}; for a detailed discussion of the method and
current measurements see \cite{Allen:07} and references therein.]

In order to study dark energy, \cite{Allen:07} use $f_{\rm gas}$
measurements for a sample of 42 hot ($kT_{2500}> 5$keV), X-ray
luminous, dynamically relaxed clusters. The $f_{\rm gas}$ measurements
are made within an angle $\theta^{\rm \Lambda CDM}_{2500}$ for each
cluster, corresponding to $r_{2500}$ for a reference flat $\Lambda$CDM
cosmology (with $\Omega_{\rm m}=0.3$ and $H_{\rm 0}=70$\kmpspMpc).
The $f_{\rm gas}$ measurements in the reference cosmology $f^{\rm
  \Lambda CDM}_{\rm gas}$ are related to the true values $f^{\rm
  true}_{\rm gas}$ as

\begin{equation}
  f^{\rm \Lambda CDM}_{\rm gas}(z;\theta^{\rm \Lambda CDM}_{2500})=f^{\rm true}_{\rm gas}(z;\theta^{\rm \Lambda CDM}_{2500})\left(\frac{d^{\rm \Lambda CDM}_{A}}{d^{\rm true}_{A}}\right)^{3/2}\,.
\label{eq:fgas}
\end{equation}

\noindent Non-radiative hydrodynamical simulations
\citep{Eke:98,Nagai:07,Crain:07} suggest that $f^{\rm true}_{\rm gas}$
is likely to be approximately constant in redshift. Thus
\citep{Allen:07},

\begin{equation}
f^{\rm true}_{\rm gas}(z;\theta^{true}_{2500})=\left(\frac{\Omega_{\rm b}}{\Omega_{\rm m}}\right)\left(\frac{b_{\rm 0}}{1+s_{\rm 0}}\right)\,,
\label{eq:uncorr}
\end{equation}

\noindent where $s_{\rm 0}=0.16h^{\rm 0.5}_{\rm 70}$
\citep{Lin:04,Gonzalez:07}\footnote{\cite{Lin:04} measured the mass in
  stars in galaxies, and included the intracluster light only as a
  model. \cite{Gonzalez:07} measured the mass in stars in both
  galaxies and intracluster light. For the largest clusters both works
  find simular results.} is the observed ratio of the mass in stars
(both in galaxies and intracluster light) to the X-ray emitting gas
mass, and $b_{\rm 0}=0.82$ \citep{Eke:98} is the depletion factor for
the baryon fraction in clusters with respect to the cosmic mean value.

As discussed by \cite{Allen:07}, an angular correction factor is also
required to account for the fact that $f^{\rm true}_{\rm
  gas}(z;\theta^{true}_{2500})$ needs not be exactly equal to $f^{\rm
  true}_{\rm gas}(z;\theta^{\rm \Lambda CDM}_{2500})$.  Observations
of large, relaxed clusters show that for the radial range of interest,
$0.7<r/r_{\rm 2500}<1.2$, the $f_{\rm gas}(r)$ profiles can be fit by
a shallow power-law model with slope
$\eta=0.214\pm0.022$.\footnote{Note that even using two very different
  reference cosmologies such as SCDM ($\Omega_{\rm m}=1$, $H_{\rm
    0}=50$\kmpspMpc) and $\Lambda$CDM ($\Omega_{\rm m}=0.3$, $H_{\rm
    0}=70$\kmpspMpc), \cite{Allen:07} obtained similar values for
  $\eta$ around $r_{\rm 2500}$.}.  Thus, we have

\begin{equation}
  f^{\rm true}_{\rm gas}(z;\theta^{\rm \Lambda CDM}_{2500})
  =f^{\rm true}_{\rm gas}(z;\theta^{\rm true}_{2500})\left(\frac{\theta^{\rm \Lambda CDM}_{2500}}{\theta^{\rm true}_{2500}}\right)^{\eta}\,,
\label{eq:angcorr}
\end{equation}

\noindent where $\theta_{\rm 2500}=r_{\rm 2500}/d_{\rm A}$, and

\begin{equation}
  \left(\frac{\theta^{\rm \Lambda CDM}_{2500}}{\theta^{\rm true}_{2500}}\right)^{\eta}=\left(\frac{[H(z)\,d_{\rm A}(z)]^{\rm true}}{[H(z)\,d_{\rm A}(z)]^{\rm \Lambda CDM}}\right)^{\eta}\,.
\label{eq:angcosm}
\end{equation}

\noindent This correction factor is small and can be neglected for most
analyses of current data, although its inclusion leads to slightly
tighter constraints on dark energy \citep{Allen:07}. However, for
future experiments of the precision being considered here, the
inclusion of the angular correction term becomes important.

\begin{table*}
\begin{center}
\caption{Systematic allowances incorporated in the $f_{\rm gas}$
and XSZ experiments.}
\label{table:sys}
\begin{tabular}{ c c c c c }
&&&&  \\   
Cluster                          & Parameter        & & Allowance (optimistic/standard/pessimistic)   &  Type    \\
\hline                                              
\noalign{\vskip 5pt}
\underline{$f_{\rm gas}$ EXPERIMENT} &&&&\\
\noalign{\vskip 4pt}
Calibration/Modelling            & $K$              & & $1.0\pm0.02/\pm0.05/\pm0.10$                  & Gaussian \\
Non-thermal pressure             & $\gamma$         & & $0.96<\gamma<1.04$/$0.92<\gamma<1.08$         & uniform  \\
Gas depletion: norm.             & $b_0$            & & $0.82\times(1\pm0.02/\pm0.05/\pm0.10)$        & uniform  \\
Gas depletion: evol. (linear)    & $\alpha_{\rm b}$ & & $\pm0.02/\pm0.05/\pm0.10$                     & uniform  \\
Gas depletion: evol. (quadratic) & $\beta_{\rm b}$  & & $\pm0.02/\pm0.05/\pm0.10$                     & uniform  \\
Stellar mass: norm.              & $s_0$            & & $0.16\times(1\pm0.02/\pm0.05/\pm0.10)$        & Gaussian \\
Stellar mass: evol. (linear)     & $\alpha_{\rm s}$ & & $\pm0.02/\pm0.05/\pm0.10$                     & uniform  \\
Stellar mass: evol. (quadratic)  & $\beta_{\rm s}$  & & $\pm0.02/\pm0.05/\pm0.10$                     & uniform  \\
\noalign{\vskip 9pt}
\underline{XSZ EXPERIMENT} &&&&\\
\noalign{\vskip 4pt}
Calibration/Modelling            & $k_0$            & & $1.0\pm0.02/\pm0.05$                          & Gaussian \\
evolution (linear)               & $\alpha_{\rm k}$ & & $\pm0.02/\pm0.05/\pm0.10$                             & uniform  \\
\noalign{\vskip 5pt}  
\hline                      
\end{tabular}
\end{center}
\end{table*}

\subsubsection{Allowances for systematic uncertainties}
\label{priors}

Following \cite{Allen:07}, we modify equation (\ref{eq:uncorr}) to
account for systematic uncertainties in the $f_{\rm gas}$ analysis:

\begin{equation}
  f^{\rm true}_{\rm gas}(z;\theta^{\rm true}_{2500})=\gamma\,K\,\left(\frac{\Omega_{\rm b}}{\Omega_{\rm m}}\right)\left(\frac{b(z)}{1+s(z)}\right).
\label{eq:fgastrue}
\end{equation}

\noindent Here $\gamma$ allows for departures from the assumption of
hydrostatic equilibrium, due to non-thermal pressure support; $K$ is a
normalization uncertainty relating to instrumental calibration and
certain modelling issues; $b(z)=b_{\rm 0}(1+\alpha_{\rm b}z+\beta_{\rm
  b}z^{\rm 2})$ accounts for uncertainties in the cluster depletion
factor, both in the normalization, $b_0$, and possible linear,
$\alpha_{\rm b}$, and quadratic, $\beta_{\rm b}$, evolution with
redshift\footnote{Note that the allowances on $\alpha_{\rm b}$ and
  $\beta_{\rm b}$ can also be assumed to encompass the combined
  uncertainties in the redshift evolution of $\gamma$, $K$ and $b$,
  which have the same effect on equation (\ref{eq:fgastrue}).};
$s(z)=s_{\rm 0}(1+\alpha_{\rm s}z+\beta_{\rm s}z^{\rm 2})$ accounts
for uncertainties in the stellar mass fraction.\footnote{Working with
  current data, \cite{Allen:07} use only the linear order of the
  redshift expansions for their systematic allowances
  i.e. $\alpha_{\rm b}$ and $\alpha_{\rm s}$.}

Using hydrodynamic N-body simulations \cite{Nagai:07} show that for
measurements at $r_{\rm 2500}$ in {\it large, relaxed clusters},
non-thermal pressure support is unlikely to exceed $8$ per
cent. Furthermore, if, as suggested by some current X-ray data
\citep{Fabian:03, Fabian:05, Reynolds:05}, the gas viscosity is higher
than that included in current simulations, then non-thermal pressure
support could be even lower. Based on these findings, we adopt by
default a uniform prior such that non-thermal pressure support lies in
the range $0-8$ per cent (although a more pessimistic range of $0-16$
per cent is also considered).  Since the use of an asymmetric prior
would bias the analysis, levering $\Omega_{\rm m}$ above the fiducial
value, we employ an equivalent, rescaled symmetric prior such that
$1-(a/2)<\gamma<1+(a/2)$, where $a=|1-1.08|/1.04$.

The depletion parameter, $b_{\rm 0}$, reflects the thermodynamic
history of the X-ray emitting cluster gas.  Using non-radiative
simulations of hot, massive clusters of comparable size to the real
clusters to be used in the $f_{\rm gas}$ experiment, \cite{Eke:98}
\cite[see also][]{Allen:04, Nagai:07, Crain:07} obtained $b_{\rm
  0}=0.82 \pm 0.03$ at the radius of the measurements $r_{\rm 2500}$
($\sim 0.25r_{\rm vir}$) and found no evidence for redshift evolution:
$\alpha_{\rm b}=0.00\pm0.03$ for measurements made at $r \sim
0.5r_{vir}$, spanning the redshift range $0<z<1$.  As discussed by
\cite{Allen:07}, however, systematic uncertainties are associated with
current predictions for $b(z)$, due to limitations in the accuracy of
the physical approximations employed in the simulations. Estimating
the residual uncertainties in the prediction of $b(z)$ that will be
appropriate at the time of a future $f_{\rm gas}$ data set ($\sim
2015-2020$) is difficult. We have chosen to use a range of values that
extend from optimistic to pessimistic scenarios (see
Table~\ref{table:sys}).

Current optical and near infrared data for low-to-intermediate
redshift clusters give $s_{\rm 0}=0.16h_{\rm 70}^{\rm -0.5}$
\citep{Fukugita:98, Lin:04, Gonzalez:07}. Although, at present, the
constraints on $s(z)$ for clusters at $z\gsim0.5$ are sparse, we
expect the form of $s(z)$ to be relatively well understood by the time
of the $f_{\rm gas}$ experiment.

In order to keep the interpretation of our results simple, we present
results for three sets of systematic allowances: for the parameters,
$K, b_{\rm 0}, \alpha_{\rm b}, \beta_{\rm b}, s_{\rm 0}, \alpha_{\rm
  s}, \beta_{\rm s}$, we employ allowances of either $\pm 2$ per cent
(optimistic), $\pm 5$ per cent (standard), or $\pm 10$ per cent
(pessimistic). In all cases, we employ uniform priors with the
exception of $K$ and $s_{\rm 0}$, for which Gaussian priors are more
appropriate and therefore used.  As noted above, a uniform allowance
of $\pm4$ per cent on $\gamma$, is included by default, although the
effects of doubling the uncertainty in this parameter are also
examined. We stress that whether $\gamma=1$ precisely, or $\alpha_{\rm
  b}, \alpha_{\rm s}$ etc are precisely zero, is not of primary
importance to a future analysis: if known, the exact values can be
incorporated into the default model. It is the uncertainties in the
values that affect the accuracy and precision of the dark energy
constraints.

\subsection{Analysis of the SZ data: the XSZ experiment}
\label{sec:sz}

For the true, underlying cosmology, the measurement of the Compton
$y-$parameter from both the X-ray and SZ data should match
\citep[e.g.][]{Molnar:02,Schmidt:04,Bonamente:06}. For a given
cosmology the y-parameter predicted by X-ray data depends on the
square root of the angular diameter distance to the cluster, $d_{\rm
  A}^{0.5}$, whereas the observed SZ flux at radio or sub-mm
wavelengths is independent of the cosmology assumed. Combining the
$y-$parameter results, we can measure the distances to the clusters as
a function of redshift and, therefore, constrain dark energy.
 
\begin{equation}
{y^{\rm \Lambda CDM}} = {y^{\rm SZobs}}\, k(z)\left( \frac{d_{\rm A}^{\rm \Lambda CDM}}{d_{\rm A}^{\rm true}}\right)^{1/2}\,.
\label{eq:ycompton}
\end{equation}

\noindent Here $y^{\rm \Lambda CDM}$ is the X-ray measurement of the
y-parameter for the reference cosmology and $y^{\rm SZobs}$ is the
radio/sub-mm observation.\footnote{The limitations of existing SZ data
  have to date restricted the XSZ experiment to measurements of the
  Hubble constant \cite[e.g.][and references therein]{Bonamente:06}.}~
Following a similar approach to that adopted with the $f_{\rm gas}$
data, we incorporate systematic allowances into equation
(\ref{eq:ycompton}): $k(z)=k_{\rm 0}(1+\alpha_{\rm k}z)$ accounts for
the combined systematic uncertainties in the X-ray and SZ data
$y-$parameter measurements due to calibration, geometric effects, gas
clumping, etc., and their evolution.  We employ Gaussian priors on
$k_{\rm 0}$ of size 2 (optimistic) or $5$ (standard/pessimistic) per
cent and uniform priors on $\alpha_{\rm k}$ of size 2 (optimistic),
$5$ (standard) or 10 (pessimistic) per cent.

We note that the best clusters to observe for the XSZ experiment are
the same systems used for the $f_{\rm gas}$ experiment: the largest,
most dynamically relaxed clusters. These are the clusters for which
the SZ signals are strongest and for which systematic uncertainties
associated with geometry and thermodynamic structure are
minimized. Note also that no additional X-ray observations are
required to carry out the XSZ experiment, once the $f_{\rm gas}$ data
are in hand.

\begin{figure*}
\includegraphics[width=3.2in]{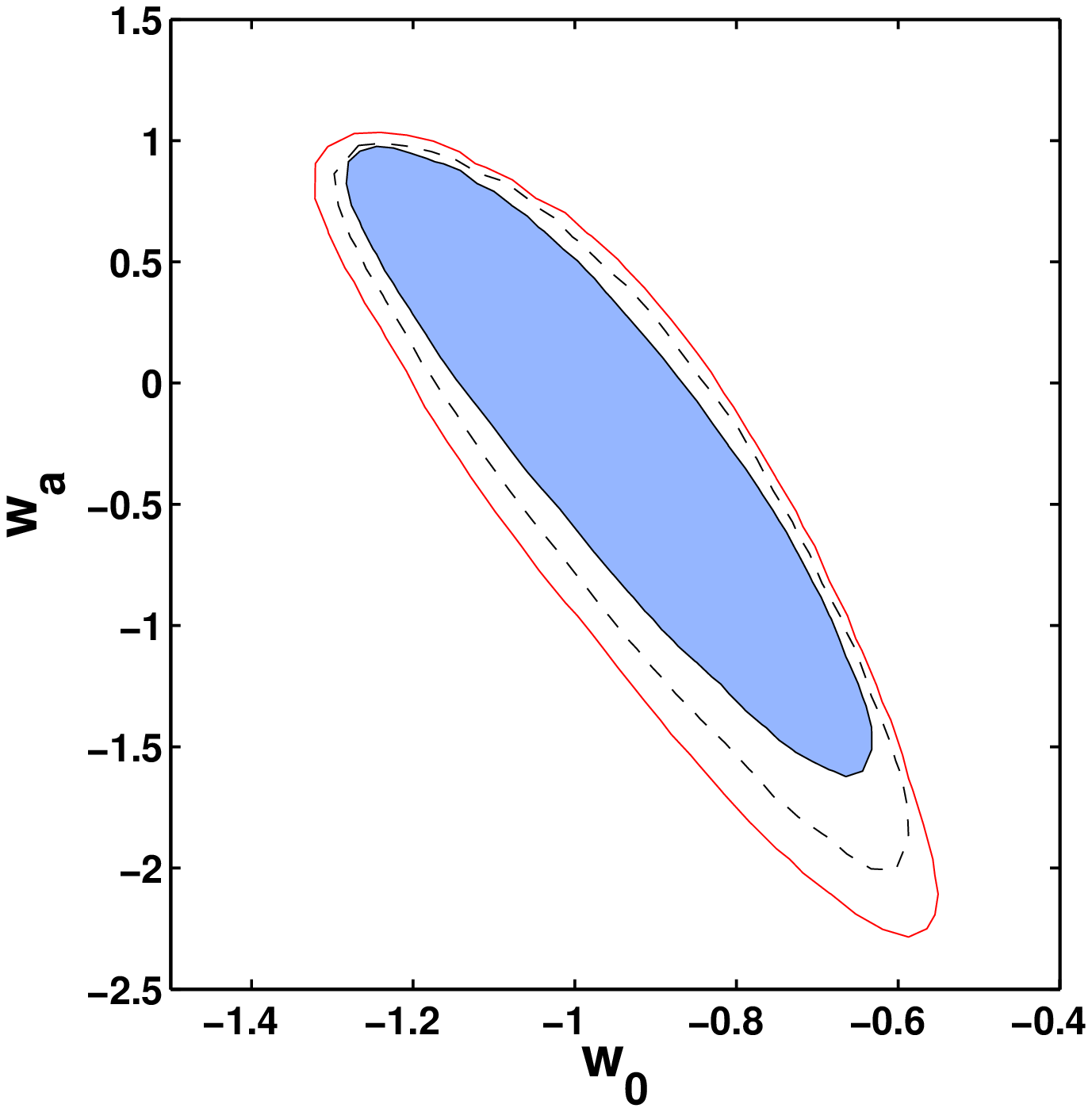}\hspace{0.6cm}
\includegraphics[width=3.2in]{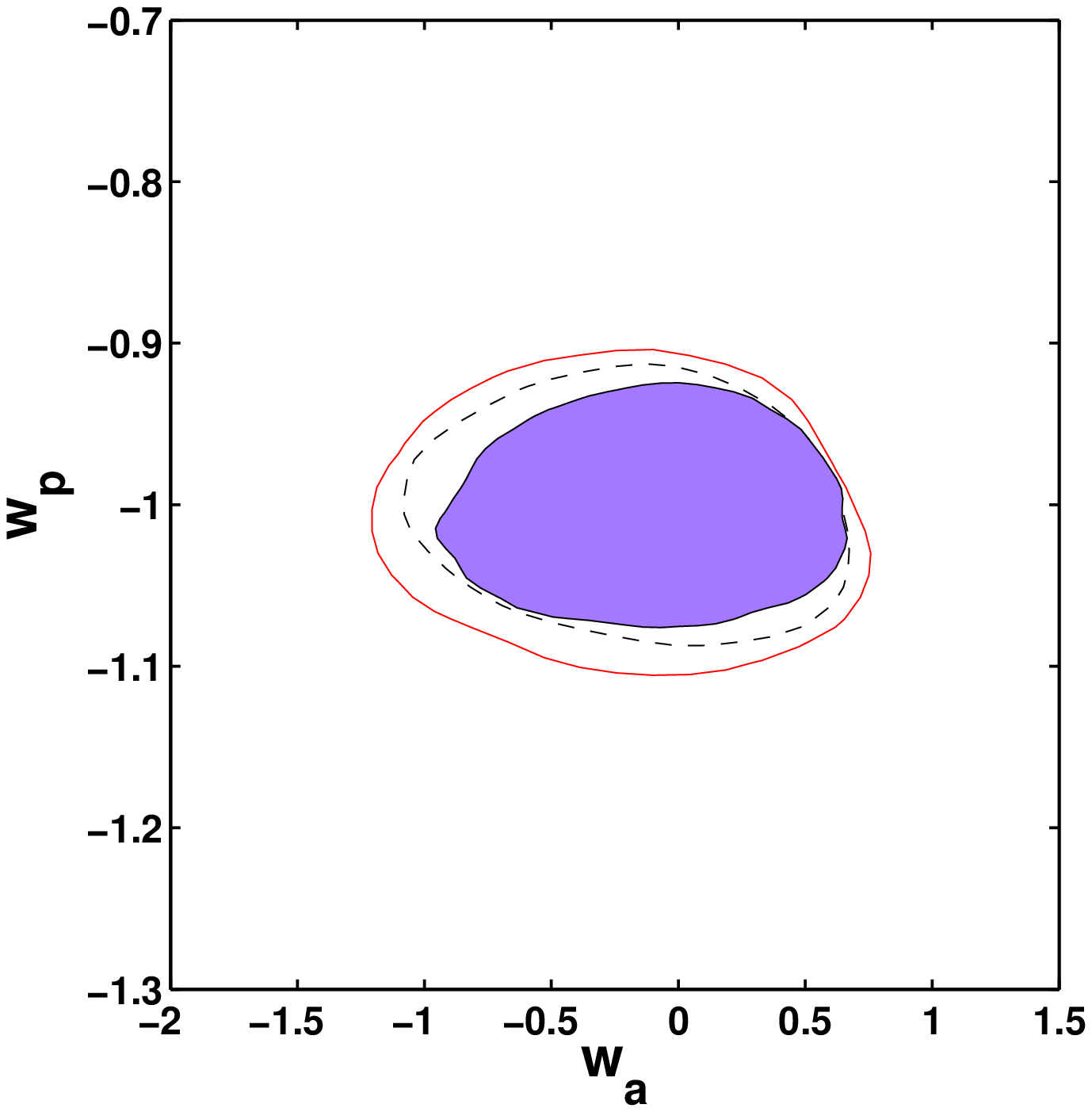}
\caption{(Left panel) The 95 per cent confidence contours in the
  $w_{\rm 0}-w_{\rm a}$ for the default dark energy model using the
  optimistic (2 per cent; blue, solid contour), standard (5 per cent;
  dashed contour) and pessimistic (10 per cent; red contour)
  allowances. (Right panel) The 68 per cent confidence contours in the
  $w_{\rm a}-w_{\rm p}$ plane for the default dark energy model using
  the optimistic (2 per cent; purple, solid contour), standard (5 per
  cent; dashed contour) and pessimistic (10 per cent; red contour)
  allowances.  The marginalized $1\sigma$ confidence intervals on
  $w_{\rm a}$ and $w_{\rm p}$ are used to calculate the FoM. The
  figure confirms that $w_{\rm a}$ and $w_{\rm p}$ are not strongly
  correlated, as assumed in the definition of the FoM (Section 2).}
\label{fig:ws}
\end{figure*}

\subsection{Incorporating the CMB data}
\label{cmbdata}

In addition to the dark energy model parameters and the $f_{\rm gas}$
parameters discussed in Section~\ref{sec:cldata}, we vary the
following eight CMB-related parameters in the MCMC analysis: the mean
physical baryon density, $\Omega_{\rm b}h^{\rm 2}$; the mean physical
cold dark matter density, $\Omega_{\rm dm}h^{\rm 2}$; the
(approximate) ratio of the sound horizon at last scattering to the
angular diameter distance \citep{kosowsky:02}, $\theta_{\rm s}$; the
optical depth to reionization (assumed to occur in a sharp
transition), $\tau$; the mean curvature density of the Universe,
$\Omega_{\rm k}$; the scalar adiabatic spectral index, $n_{\rm s}$;
and the scalar adiabatic amplitude, $A_{\rm s}$, at $k=0.05
\hbox{$\Mpc^{-1}$}$. We employ a uniform prior on $\ln(A_{\rm
  s})$. The combination of $\theta_{\rm s}$ and $\ln(A_{\rm s})$ as
parameters, rather than $H_{\rm 0}$ and $A_{\rm s}$, leads to a more
Gaussian probability density distribution which, in turn, aids
sampling \citep{kosowsky:02, Lewis:06}.

The degeneracies between dark energy model parameters and $\Omega_{\rm
k}$ are of particular importance in the analysis
\citep{Rapetti:05,Clarkson:07}.\footnote{\cite{Allen:04,Allen:07} and
\cite{Rapetti:05} showed that the combination of $f_{\rm gas}$ plus
CMB data allows one to drop both the assumption of flatness and the
priors on $\Omega_{\rm b}h^{\rm 2}$ and $h$ that would otherwise be
required for the $f_{\rm gas}$ analysis. The $f_{\rm gas}$+CMB data
combination also alleviates other important parameter degeneracies,
e.g. between $\Omega_{\rm b}h^{\rm 2}$, $n_{\rm s}$ and $\tau$.} For
their forecasts, the DETF include {\sc Planck} priors in their Fisher
matrix analysis, approximating the role of future CMB constraints 
as well as the degeneracies between the dark energy parameters and
$\Omega_{\rm k}$. Here, we account fully for the degeneracies 
between parameters, and the complementarity of the data sets.

Given the vector, ${\bf \epsilon}$, of CMB-related parameters, we
sample the exponential of the following mean logarithmic likelihood
\citep{Lewis:05, Lewis:06}

\begin{equation}
  \langle\ln P({\bf \epsilon}|{\bf \epsilon}_{\rm 0})\rangle=-\frac{1}{2}[Tr(C_{\rm l}({\bf \epsilon}_{\rm 0})C_{\rm l}({\bf \epsilon})^{\rm -1})+\ln|C_{\rm l}({\bf \epsilon})|]\,,
\label{eq:logli}
\end{equation}

\noindent where ${\bf \epsilon}_{\rm 0}$ is the vector formed by the
corresponding fiducial values of Table~\ref{tab:fidu}. Note that where
the posterior is non-Gaussian, the marginalized constraints on
individual parameters need not peak exactly at the fiducial values,
although for all cases considered here the differences are very small.

\begin{table*}
\begin{center}
  \caption{The $1\sigma$ uncertainties on the dark energy parameters
    and FoM. Systematic allowances of 2 per cent (optimistic), 5 per
    cent (standard) or 10 per cent (pessimistic) have been used.
    Results are presented for the default model and for six other
    cases, described in the text. We obtain a FoM in the range
    $34-43$, for the optimistic allowances, in the range $21-33$, for
    the standard, and in the range $15-29$, for the pessimistic.}

\label{tab:models}
\begin{tabular}{ c c c c c c c c c }
\hline
\multicolumn{2}{c}{Run} &
\multicolumn{5}{c}{Dark energy parameters} &
\multicolumn{1}{c}{FoM} &
\multicolumn{1}{c}{$\Delta$FoM/FoM} \\


Allowances & Model & $\hat{\sigma}(\Omega_{\rm m})$ & $\hat{\sigma}(\Omega_{\rm de})$ & $\hat{\sigma}(w_{\rm 0})$ & $\hat{\sigma}(w_{\rm p})$ & $\hat{\sigma}(w_{\rm a})$ & $[\hat{\sigma}(w_{\rm p})\times\hat{\sigma}(w_{\rm a})]^{-1}$ & (percentage) \\
\hline

\noalign{\vskip 5pt}

$ 2\% $ & Default & 0.014 & 0.011 & 0.130 & 0.050 & 0.52 & 38.5 & -- \\
\noalign{\vskip 5pt}

$ 2\% $ & DE clustering & 0.014 & 0.010 & 0.137 & 0.048 & 0.56 & 37.2 & -3.4\% \\
\noalign{\vskip 5pt}

$ 2\% $ & CMB conservative & 0.015 & 0.011 & 0.135 & 0.052 & 0.57 & 33.7 & -12.5\% \\
\noalign{\vskip 5pt}

$ 2\% $ & Quadratic & 0.014 & 0.010 & 0.146 & 0.050 & 0.56 & 35.7 & -7.3\% \\
\noalign{\vskip 5pt}

$ 2\% $ & Half sample & 0.014 & 0.011 & 0.125 & 0.049 & 0.51 & 40.0 & +3.9\% \\
\noalign{\vskip 5pt}

$ 2\% $ & Double $\gamma$ & 0.019 & 0.013 & 0.129 & 0.055 & 0.52 & 35.0 & -9.1\% \\
\noalign{\vskip 5pt}

$ 2\% $ & Adding XSZ & 0.013 & 0.010 & 0.122 & 0.047 & 0.50 & 42.6 & +10.6\% \\
\noalign{\vskip 5pt}

\noalign{\vskip 8pt}

$ 5\% $ & Default & 0.022 & 0.016 & 0.137 & 0.056 & 0.58 & 30.8 & -- \\
\noalign{\vskip 5pt}

$ 5\% $ & DE clustering & 0.020 & 0.016 & 0.152 & 0.055 & 0.69 & 26.4 & -14.3\% \\
\noalign{\vskip 5pt}

$ 5\% $ & CMB conservative & 0.023 & 0.017 & 0.147 & 0.058 & 0.66 & 26.1 & -15.3\% \\
\noalign{\vskip 5pt}

$ 5\% $ & Quadratic & 0.022 & 0.016 & 0.157 & 0.079 & 0.61 & 20.8 & -32.5\% \\
\noalign{\vskip 5pt}

$ 5\% $ & Half sample & 0.022 & 0.016 & 0.132 & 0.058 & 0.57 & 30.2 & -1.9\% \\
\noalign{\vskip 5pt}

$ 5\% $ & Double $\gamma$ & 0.024 & 0.018 & 0.137 & 0.058 & 0.58 & 29.7 & -3.6\% \\
\noalign{\vskip 5pt}

$ 5\% $ & Adding XSZ & 0.021 & 0.015 & 0.132 & 0.053 & 0.57 & 33.1 & +7.5\% \\
\noalign{\vskip 5pt}

\noalign{\vskip 8pt}

$ 10\% $ & Default & 0.033 & 0.024 & 0.143 & 0.064 & 0.62 & 25.2 & -- \\
\noalign{\vskip 5pt}

$ 10\% $ & DE clustering & 0.029 & 0.022 & 0.164 & 0.059 & 0.73 & 23.2 & -7.9\% \\
\noalign{\vskip 5pt}

$ 10\% $ & CMB conservative & 0.038 & 0.027 & 0.163 & 0.069 & 0.73 & 19.9 & -21.0\% \\
\noalign{\vskip 5pt}

$ 10\% $ & Quadratic & 0.033 & 0.025 & 0.173 & 0.106 & 0.64 & 14.7 & -41.7\% \\
\noalign{\vskip 5pt}

$ 10\% $ & Half sample & 0.033 & 0.024 & 0.137 & 0.064 & 0.60 & 26.0 & +3.2\% \\
\noalign{\vskip 5pt}

$ 10\% $ & Double $\gamma$ & 0.034 & 0.025 & 0.144 & 0.065 & 0.61 & 25.2 & 0.0\% \\
\noalign{\vskip 5pt}

$ 10\% $ & Adding XSZ & 0.028 & 0.020 & 0.134 & 0.059 & 0.59 & 28.7 & +13.9\% \\
\noalign{\vskip 5pt}

\hline
\end{tabular}
\end{center}
\end{table*} 

\subsection{Dark energy clustering}
\label{sec:depert}

For a dark energy model with a constant equation of state,
$w$, \cite{Weller:03} and \cite{Bean:03} showed that dark energy
clustering can have a non-negligible impact on the constraints, driven
primarily by the effect of such perturbations on the Integrated
Sachs-Wolfe (ISW) effect. \cite{Spergel:06} showed that accounting for
dark energy clustering has a large effect on current dark energy
constraints derived from CMB data alone. Since, for constant$-w$
models, combining the CMB data with e.g. distance measurements from
type Ia supernovae or X-ray galaxy clusters leads to tight constraints
on $w$ and a result consistent with a cosmological constant ($w=-1$,
for which no dark energy clustering occurs), the importance of
accounting for dark energy perturbations is reduced \citep{Weller:03,
Rapetti:05, Spergel:06}. However, when one considers more general
models in which $w$ evolves, \cite{Rapetti:05} showed that even with
the best current data combinations, accounting for the effects of dark
energy perturbations is important: the constraints on $w(a)$ increase
by a factor of $\sim 2$ with respect to the case where dark energy
clustering is (wrongly) ignored.

For our analysis, we assume that dark energy is an imperfect fluid
where dissipative processes generate entropy perturbations. As
suggested by quintessence scenarios, we assume a constant, general
(non-adiabatic) sound speed $\hat{c}_{\rm s}^{\rm 2}=1$ in the
comoving frame of the fluid (denoted by the circumflex $\hat{}$~).
This is the only frame for which the general sound speed is gauge
invariant \citep{Bean:03}. Following \cite{Weller:03} and
\cite{Bean:03}, \cite{Rapetti:05} extended the dark energy
perturbation equations to account for an evolving dark energy equation
of state, $w(a)$. As in \cite{Rapetti:05} we calculate the density,
$\delta$, and velocity, $v$, perturbation equations \citep{Ma:95} in
the synchronous gauge

\begin{eqnarray}
\dot{\delta}&=&-3\mathcal{H}(\hat{c}_{\rm s}^{\rm 2}-w)\hat{\delta}-(1+w)(kv+3\dot{\mathcal{B}})+\mathcal{E}(\dot{w})\label{eq:densitypert} \\
\dot{v}&=&-\mathcal{H}(1-3\hat{c}_{\rm s}^{\rm 2})v+\frac{k\hat{c}_{\rm s}^{\rm 2}\delta}{1+w} \,,
\label{eq:velocitypert}
\end{eqnarray}

\noindent where both derivatives, denoted by dots, and the Hubble
parameter, $\mathcal{H}$, are with respect to conformal time.
$\mathcal{B}=\delta a/a$ is the metric perturbation and $\hat{\delta}$
is the density perturbation in the comoving frame of the dark energy
fluid. The density perturbation, $\hat{\delta}$, can be recast into
the CDM comoving frame density and velocity perturbations, $\delta$
and $v$, using the relation given by \cite{Kodama:84}, $\hat{\delta} =
\delta+3\mathcal{H}(1+w)v/k$. Using this relation in equation
(\ref{eq:densitypert}), we recover the perturbation equations of
\cite{Weller:03,Bean:03} except that here $w$ depends on the scale
factor $a$ and introduces a new source term, $\mathcal{E}(\dot{w})=
3\mathcal{H}\dot{w}v/k$, in the density perturbation equation. Note
that this term depends on the derivative of the equation of state,
$\dot{w}$. Equation (\ref{eq:velocitypert}) does not have a new
term.

As discussed by \cite{Vikman:05, Caldwell:05} a single classical
scalar field cannot evolve from a quintessence-like, $w>-1$, to
phantom-like, $w<-1$, behavior. However, \cite{Onemli:02, Onemli:04}
proposed a single scalar field model where a super-accelerated phase
($w<-1$) of the cosmic expansion can be achieved via quantum effects.
Later, \cite{Kahya:06} showed that this model is stable.
Alternatively, \cite{Feng:05, Guo:05, Hu:05, Zhao:05} suggested the
so-called quintom model: in this model, the effective equation of
state of two combined scalar fields, one with $w>-1$ and the other
with $w<-1$, can cross the cosmological-constant boundary, $w=-1$, as
it evolves in time. Other models that allow $w(a)$ to cross this
boundary have also been proposed. However, for practical purposes,
using an effective, evolving dark energy equation of state produces a
well-known divergence in equation (\ref{eq:velocitypert}) when
$w(a)=-1$. This divergence can be avoided \citep{Huey:04, Caldwell:05}
by imposing $\dot{\delta}=0$ and $\dot{v}=0$ within the logarithmic
singularity region, $w=-1+|\epsilon|$, where $\epsilon$ is
infinitesimally small.  Inaccuracies in this approximation have a
negligible impact on the resulting CMB power spectra
\citep{Rapetti:05, Xia:05, Xia:07}.

In what follows, we present results for cases where dark energy
clustering is either accounted for or ignored in the CMB analysis
(Section~\ref{deevo}). Dark energy clustering does
not have a significant impact on the $f_{\rm gas}$ analysis.

\section{Results}
\label{constraints}

As described in Section~\ref{sec:de}, to enable a direct comparison
with the predicted dark energy constraints for other planned
experiments, we parameterize our results in terms of the DETF FoM. We
present results for a fiducial $f_{\rm gas}$+CMB data set,
incorporating the statistical uncertainties and systematic allowances
described above, and with zero scatter about the fiducial curves. The
absence of scatter in the simulated samples ensures that the peaks of
the posterior probability distributions occur at the expected values,
in the same way that the DETF Fisher matrix analysis does. That is, in
order to compare our results with the DETF, we select the same
realization as the DETF. Note also that the use of a zero-scatter
realization does not affect the FoM. We have explicitly confirmed
this, and that using other realizations does not have a large impact
on the FoM, by comparing the FoM for the fiducial realization to a
series of Monte Carlo simulations, in which appropiate scatter about
the fiducial $f_{\rm gas}$ curve was included.

\subsection{Constraints on the FoM}
\label{sec:sce}

\begin{figure}
\includegraphics[width=3.2in]{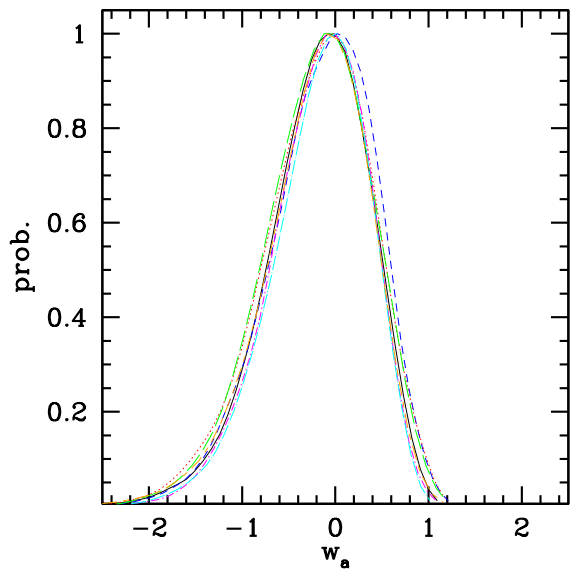}
\caption{Posterior probability distributions for $w_{\rm a}$ for the
  cases described in Table~\ref{tab:models}: default scenario (black,
  solid line), with dark energy clustering (blue, long dashed line),
  conservative CMB data (red, dotted line), including quadratic
  evolution allowances (green, long dashed line), using the
  250-cluster sample with $3.5$ per cent measurement errors (magenta,
  dot-dashed line), doubling the systematic allowance on $\gamma$
  (orange, long-short dashed) and adding the XSZ experiment (cyan,
  long dot-dashed line). The (optimistic) 2 per cent systematic
  allowances are used in every case.}
\label{fig:wdistr}
\end{figure}

In the first case, we determine constraints for our `default'
analysis: this involves $f_{\rm gas}$ data for 500 clusters measured
to $5$ per cent accuracy and CMB data with negligible foreground
contamination.  We ignore the effects of dark energy clustering (as do
the DETF) and allow linear evolution in $b(z)$ and $s(z)$. No
follow-up SZ data are included.

The constraints from the default analysis are shown in
Figures~\ref{fig:ws},~\ref{fig:fom}, and ~\ref{fig:evol}. The left
panel of Figure~\ref{fig:ws} shows the well-known degeneracy between
$w_{\rm 0}$ and $w_{\rm a}$. The right panel of that figure shows the
constraints in the $w_{\rm a}-w_{\rm p}$ plane. The results from the
MCMC analysis confirm that $w_{\rm p}$ and $w_{\rm a}$ are
approximately uncorrelated, which facilitates the simple calculation
of the FoM=$[\hat{\sigma}(w_{\rm p})\times\hat{\sigma}(w_{\rm
  a})]^{-1}$ a described in Section~\ref{sec:de}.

Table~\ref{tab:models} summarizes the results on $\Omega_{\rm m}$,
$\Omega_{\rm de}$, $w_{\rm 0}$, $w_{\rm a}$, $w_{\rm p}$ and the FoM
for the default analysis and 2, 5 or 10 per cent systematic
allowances.  Also included in the table are results for five further
slightly modified, interesting scenarios: for the case where we
include dark energy clustering (Section~\ref{sec:depert}); for the
case where we use the more conservative CMB data set
(Section~\ref{planckdata}); for the case where we allow quadratic
redshift evolution in the gas depletion factor, $\beta_{\rm b}$, and
the stellar fraction, $\beta_{\rm s}$ (Section~\ref{priors}); for the
case we use the 250-cluster $f_{\rm gas}$ data set with $3.5$ per cent
measurement errors; for the case where we double the systematic
allowance on $\gamma$; and for the case where we include extra
information from the XSZ experiment (Section~\ref{sec:sz}). For each
of these scenarios, we list results for 2 per cent (optimistic), 5 per
cent (standard) and 10 per cent (pessimistic) allowances.

Interestingly, we see that constraints on dark energy are similar for
most cases of interest (this is also shown graphically in
Figure~\ref{fig:wdistr}). With the optimistic, 2 per cent systematic
allowances, a FOM in the range $34-43$ is obtained. For the standard,
5 per cent systematic tolerances, the FOM lies in the range $21-33$.
Even with the pessimistic 10 per cent systematic allowances,
we obtain a FoM in the range $15-29$.

The final column of Table~\ref{tab:models} summarizes the percentage
differences in the FoM with respect to the default model for each case
of interest. We see that, using the standard or pessimistic (5 and 10
per cent allowances) the greatest impact on the FoM occurs by allowing
quadratic evolution in the systematic allowances; in this case a $\sim
30-40$ per cent reduction in the FoM with respect to the default model
is observed. Using the standard, 5 per cent allowances, we see that
accounting for dark energy clustering has only a small effect ($\sim
15$ per cent); the inclusion of the XSZ data only leads to a modest
improvement in the FoM ($\sim 10$ per cent). Doubling the uncertainty
on $\gamma$ does not have a major effect on the results, and for the
pessimistic scenario, becomes negligible.

\subsection{Comparison with DETF results}

Comparing our results on the FoM with those reported by the DETF
\citep[page 77 of the DETF report;][]{Albrecht:06}, we find that the
$f_{\rm gas}$ experiment~\footnote{The $f_{\rm gas}$ experiment
  described here would fall under the category of stage IV
  experiments, as defined by the DETF.}  has similar dark energy
constraining power to other leading, future (DETF stage IV) ground or
space-based experiments.

\begin{figure}
\includegraphics[width=3.2in]{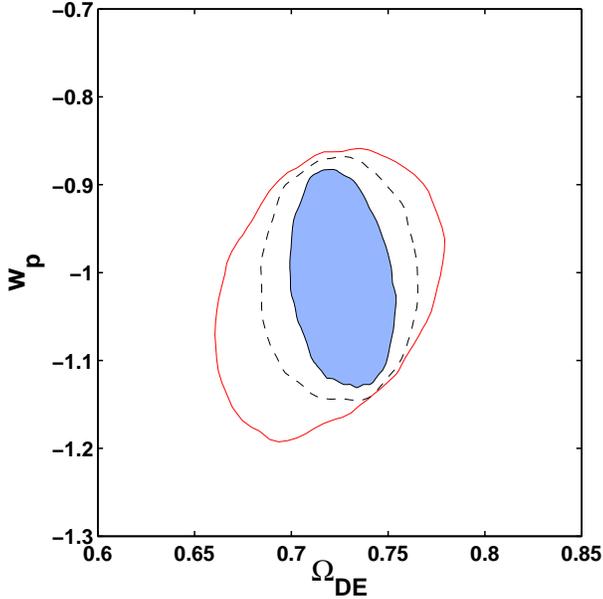}
\caption{The 95 per cent confidence contours in the $\Omega_{\rm
    de}-w_{\rm p}$ plane for the default dark energy model and
  optimistic (2 per cent; blue, solid contour), standard (5 per cent;
  dashed contour) and pessimistic (10 per cent; red contour)
  allowances. The axes are scaled to cover the same region as the
  figures presented by the DETF.}
\label{fig:fom}
\end{figure}

Figure~\ref{fig:fom} shows the 95 per cent confidence constraints in
the $\Omega_{\rm de}-w_{\rm p}$ plane for the default model and
optimistic (2 per cent), standard (5 per cent) and pessimistic (10 per
cent) allowances. The size of this confidence region is inversely
proportional to the FoM. The DETF \citep{Albrecht:06} present similar
figures, with the same axis scaling, for the other, future dark energy
experiments.  The comparable constraining power and complementary
nature of the $f_{\rm gas}$ and other experiments can (at least in
part) be seen by comparing these figures. In particular, the power of
the $f_{\rm gas}$(+Planck) experiment in constraining $\Omega_{\rm
  de}$ is evident.

\subsection{The relevance of the redshift distribution}
\label{redshift_distr}

To calculate the dark energy results presented in
Table~\ref{tab:models}, we employ a redshift distribution of clusters
drawn from a simulated X-ray luminosity function \cite[based on the
work of][as discussed in Section~\ref{sec:lf}]{Mantz:07} for both a
given fiducial cosmology and a future, planned X-ray cluster
survey. As discussed in Section~\ref{sec:selec}, we select clusters
from this distribution using the same criterion ($kT>5$keV) than we
use for current data \citep{Allen:07}. To apply this selection
criterion to the simulated data set we use a luminosity-temperature
relation obtained from present-day data \citep{Reiprich:02}. Note that
in this relation the temperatures are emission weighted instead of
mass weighted, as they are in \cite{Allen:07}, which makes our
selection even more conservative.

Finally, we select only relaxed clusters by scaling the distribution
with a factor $1/8$ (or $1/16$ for the half sample scenario). This
factor is especially conservative at low redshifts ($z\lsim 0.5$),
where the MACS survey has shown that $1/4$ clusters are sufficiently
relaxed for $f_{\rm gas}$ work. This suggests that selecting a
distribution with more clusters at low redshifts is a plausible
alternative. Furthermore, after all these conservative cuts we obtain
a total number of clusters suitable for $f_{\rm gas}$ work larger than
$500$. This offers us additional freedom to build an alternative
redshift distribution. In this section, we use two alternative
distributions to assess the impact that choosing a particular
distribution has on the dark energy constraints.

Within the limits of the analysis described above, we design two test
distributions such that their peaks are at a lower redshift, $z \sim
0.5$, than that of the original distribution, $z \sim 0.65$. From each
test distribution we form an $f_{\rm gas}$ sample of $\sim 500$
clusters, and $\sim 5$ per cent $f_{\rm gas}$ measurement errors per
cluster, as we did for the original distribution. The first test
distribution is approximately gaussian, with a large number of
clusters at low redshift, a small tail at high redshifts, and very few
clusters beyond redshift $1$. The second test distribution is similar
to the original, but shifted towards lower redshifts. At high
redshifts this distribution has less clusters than the original but
significantly more clusters than the first test distribution.

Using our default dark energy model, and the $2$ per cent set of
systematic allowances, the first test distribution provides an
increase in the FoM of $\sim 10$ per cent with respect to the original
distribution. As mentioned in Section~\ref{sec:fgas_distr}, the pivot
redshift, $z_{\rm p}$, is important for the DETF FoM
criterion. Figure~\ref{fig:evol} shows that for the $f_{\rm gas}$
experiment, $z_{\rm p}\sim 0.25$. Interestingly, the second test
distribution provides an increase in the FoM of $\sim 40$ per cent,
which suggests that having enough high redshift clusters is also
important for the DETF FoM. Note also that to obtain the $f_{\rm gas}$
measurements for each of the test samples, we will require a shorter
total exposure time than that estimated for the original sample ($\sim
15$Ms). Thus, using the same exposure time, an even larger increase in
the FoM should be achievable with such samples.

These results indicate that a further analysis is required to
determine the optimal redshift distribution of clusters with which to
carry out future $f_{\rm gas}$ experiments. Such analysis is beyond
the scope of this paper, but we will pursue it in a forthcoming
publication.

\subsection{The evolution of dark energy}
\label{deevo}

\begin{figure}
\includegraphics[width=3.2in]{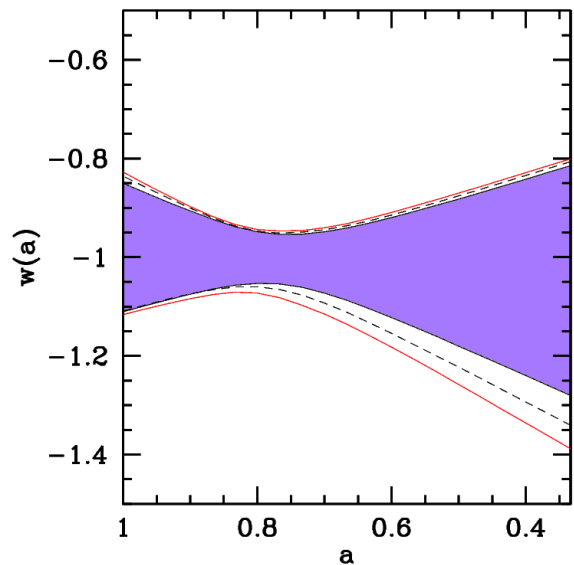}
\caption{The 1$\sigma$ confidence constraints on the evolution of the
  dark energy equation of state as a function of scale factor $w(a)$.
  Results are shown for the default model (Table~\ref{tab:spec}) using
  the optimistic (2 per cent; shaded, purple region), standard (5 per
  cent; dashed line) and pessimistic (10 per cent; solid, red line)
  systematic allowances. The tightest constraints on $w(a)$ occur at
  the pivot scale factor, $a_{\rm p}\sim 0.8$ ($z_{\rm p}\sim 0.25$).}
\label{fig:evol}
\end{figure}

Figure~\ref{fig:evol} shows the evolution of the dark energy equation
of state as a function of scale factor, $w(a)$.  The pivot scale
factor, $a_{\rm p}$, is the scale factor at which we obtain the
tightest constraint on $w$ [that measurement being $\sigma(w_{\rm
  p})]$. As shown in Figure~\ref{fig:evol}, for the $f_{\rm gas}$
experiment we measure a pivot scale factor of $a_{\rm p}\sim 0.8$,
which corresponds to a pivot redshift $z_{\rm p}\sim 0.25$.
Interestingly, these values lie between the pivot scale
factors/redshifts reported by the DETF for SNIa experiments ($a_{\rm
  p}\sim 0.93$/$z_{\rm p}\sim 0.075$) and galaxy cluster number
counts, weak lensing and BAO experiments ($a_{\rm p}\sim 0.65$/$z_{\rm
  p}\sim 0.54$).  Pinning down the evolution of $w$ over a wide
redshift range will be a crucial for unraveling the nature of dark
energy. Our results argue that a combination of $f_{\rm gas}$, CMB,
BAO, SNIa, weak lensing and galaxy cluster number count experiments is
likely to prove powerful in this regard.

\subsection{The CMB data and early dark energy}
\label{cmbede}

The acoustic scale at last scattering, $l_{\rm a}$, is tightly
constrained by CMB data \citep{Page:03} and is highly sensitive to the
amount of dark energy at recombination. $l_{\rm a}$ changes
drastically if the early dark energy density exceeds the matter plus
radiation density \citep{Wright:07}. CMB constraints on $l_{\rm a}$
provide a strong constraint on dark energy parameters at early times
\citep{Doran:01}, defining a well-known boundary in the $w_{\rm
  0}-w_{\rm a}$ plane \citep{Rapetti:05, Upadhye:05,
  Wright:07}.\footnote{The presence of the boundary in the $w_{\rm
    0}-w_{\rm a}$ plane \citep{Rapetti:05, Upadhye:05, Wright:07}
  makes it important to consider, as we do here, simulations that
  account fully for measurement uncertainties but which do not scatter
  about the fiducial curve.  Otherwise, scatter towards the CMB
  boundary would increase the FoM, and scatter away would decrease it,
  complicating the interpretation of results.} At late times, for our
experiment, dark energy is constrained primarily by the $f_{\rm gas}$
data, with a small contribution from the Integrated Sachs-Wolfe (ISW)
effect in the CMB.

\begin{figure}
\includegraphics[width=3.2in]{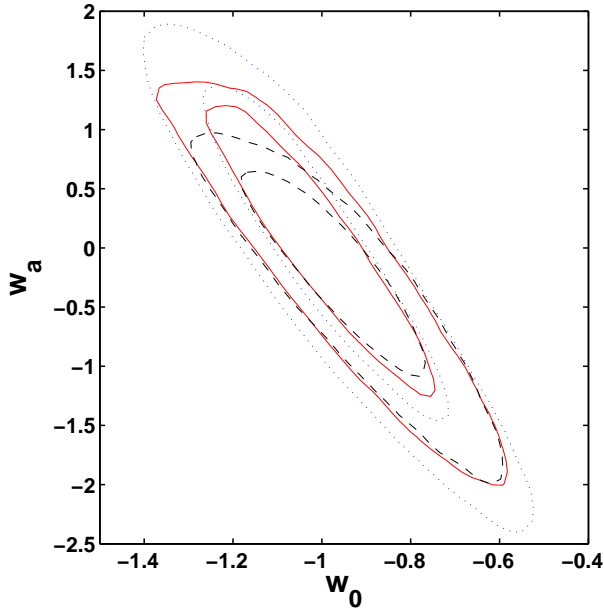}
\caption{The 68 and 95 per cent confidence contours in the $w_{\rm
0}-w_{\rm a}$ plane determined from the $f_{\rm gas}$+CMB data (black,
dashed contours) using the default dark energy model and $5$ per cent
systematic allowances.  The solid red lines show the constraints
obtained from the $f_{\rm gas}$ data alone, using priors on
$\Omega_{\rm b}h^2$, $\Omega_{\rm dm}h^2$, $l_{\rm a}$ and $h$, as
described in the text (Section~\ref{cmbede}).  The blue, dotted lines
show the constraints from the $f_{\rm gas}$ alone using priors on
$\Omega_{\rm b}h^2$ and $h$ and assuming flatness. The figure shows
how the CMB data contribute in constraining dark energy, especially
at early times.}
\label{fig:cmbede}
\end{figure}

A simple exercise provides further insight into how the CMB data help
in constraining dark energy. For this, we re-examine the constraints
in the $w_{\rm 0}-w_{\rm a}$ plane obtained from the $f_{\rm gas}$+CMB
data; the 68 and 95 per cent confidence contours for the default model
with 5 per cent allowances are shown (dashed curves) in
Figure~\ref{fig:cmbede}.  The combination of $f_{\rm gas}$+CMB data
provides tight constraints on $\Omega_{\rm b}h^2$, $\Omega_{\rm
  dm}h^2$ and $l_{\rm a}$ (driven primarily by the CMB data), and on
$h$ (driven by the combination of both data sets).  Using these
constraints as priors, we examine the constraints in the $w_{\rm
  0}-w_{\rm a}$ plane that can be obtained from the $f_{\rm gas}$ data
alone; the results are shown as the red, solid curves in
Figure~\ref{fig:cmbede}.  We see that the priors encompass some of the
CMB constraining power, in particular in defining the characteristic
upper boundary in the $w_{\rm 0}-w_{\rm a}$ plane. However, they do
not contain the full information on, e.g., the covariance of
$\Omega_{\rm b}h^2$, $\Omega_{\rm dm}h^2$ and $h$
\citep{Rapetti:05,Wright:07} which is also important in constraining
dark energy at later times.

We note that the prior on $l_{\rm a}$ provides a tight constraint on
the curvature. The blue dotted curves in Figure~\ref{fig:cmbede} show
the constraints obtained from the $f_{\rm gas}$ data alone, using only
the priors on $\Omega_{\rm b}h^2$ and $h$ and assuming flatness.

\subsection{The importance of the XSZ experiment}
\label{sec:szres}

The XSZ technique provides a complementary and independent experiment
to measure dark energy.  Although the inclusion of constraints from
the XSZ experiment leads to only modest formal improvements in the FoM
with respect to the results for the $f_{\rm gas}+$CMB data
(Table~\ref{tab:models}; as can be expected given the relatively
weak dependence on dark energy in equation 21), it is important to note that the 
XSZ experiment relies on different assumptions
and has different systematic uncertainties.  In particular, the
XSZ experiment is independent of assumptions regarding hydrostatic
equilibrium, the depletion factor, and the stellar mass fraction.
Thus, the combination of data from the $f_{\rm gas}$ and XSZ
techniques can help to ensure robustness in the results. In principle,
the inclusion of XSZ data can also allow some of the priors
in the $f_{\rm gas}$ experiment to be relaxed.

\section{Conclusions}
\label{conclusions}

We have examined the ability of a future X-ray observatory, with
capabilities similar to those planned for Constellation-X, to
constrain dark energy via the $f_{\rm gas}$ experiment. We find that
$f_{\rm gas}$ measurements for a sample of 500 hot
($kT_{2500}\gsim5$keV), X-ray bright, dynamically relaxed clusters,
with a precision of $\sim 5$ per cent, can be used to constrain dark
energy with a FoM of $15-40$. These constraints are comparable to
those predicted by the DETF \citep{Albrecht:06} for other leading,
planned (DETF Stage IV) dark energy experiments. We also find that,
for the $f_{\rm gas}$ experiment, the FoM can be boosted up by at
least $\sim 40$ per cent by selecting an optimal redshift distribution
of suitable clusters on which to carry out the $f_{\rm gas}$
observations. Interestingly, the optimal redshift distribution of
$f_{\rm gas}$ measurments appears to be shifted towards low redshifts.

As discussed in the text, a future $f_{\rm gas}$ experiment will need
to be preceded by a large X-ray or SZ cluster survey that will find
hot, X-ray luminous clusters out to high redshifts. A survey such as
that planned with the Spectrum-RG/eROSITA mission should find several
thousand of such clusters. Short `snapshot' follow-up observations of
the clusters with a new, large X-ray observatory should be able to
identify a sample of $\sim 500$ suitable systems for $f_{\rm gas}$
work. Attaining a precision of $\sim 5$ per cent with individual
$f_{\rm gas}$ measurements should be straightforward for an
observatory with characteristics similar to Constellation-X, requiring
exposure times of $\sim 20$ks on average. We note that the population
of galaxy clusters in the redshift, temperature and X-ray luminosity
range of interest has already been partially probed by the MACS survey
\citep{Ebeling:01}; Chandra observations of MACS clusters are used
extensively in current $f_{\rm gas}$ studies
\citep{Allen:04,LaRoque:06,Allen:07}. The low-level of X-ray flux
contamination from point sources observed in MACS clusters also
alleviates the requirements on the instrumental PSF for dark energy
work via the $f_{\rm gas}$ method.

In determining the predicted dark energy constraints, we have employed
the same MCMC method used to analyze current data. The MCMC method
encapsulates all of the relevant degeneracies between parameters and
allows one to easily and efficiently incorporate priors and allowances
in the analysis. We have included an array of such systematic
allowances, with tolerances ranging from optimistic to
pessimistic. Our technique differs from the DETF \citep{Albrecht:06},
who use a simpler Fisher matrix approach in the prediction of dark
energy constraints. Despite these differences, we have endeavored to
make our calculations of the FoM (Section 2) as comparable as
possible.

Benchmarking our results against those of the DETF for other, future
`Stage IV' dark energy experiments i.e. large, long-term missions, we
find that the $f_{\rm gas}$ experiment should provide a comparable FoM
to future ground-based SNIa (FoM=$8-22$), space-based SNIa
(FoM=$19-27$), ground-based BAO (FoM=$5-55$), space-based BAO
(FoM=$20-42$) and space-based cluster counting (FoM=$6-39$)
experiments. Formally, the predicted FoM for the $f_{\rm gas}$
experiment is comparable to `pessimistic' scenarios for weak lensing
experiments discussed by \cite{Albrecht:06}, although the value falls
short of the most optimistic DETF weak lensing predictions.  The tight
constraints on $\Omega_{\rm m}$ and $\Omega_{\rm de}$ for the $f_{\rm
  gas}$ experiment will be of importance when used in combination with
other techniques. Interestingly, the `pivot point' for the $f_{\rm
  gas}$ experiment lies between those of the SNIa and BAO/weak
lensing/cluster number count experiments, offering excellent redshift
coverage in attempts to pin down the evolution of dark energy.

We conclude that the $f_{\rm gas}$ experiment offers a powerful
approach for dark energy work, which should be competitive with and
complementary to the best other planned dark energy experiments.

\section*{Acknowledgments}

We thank the members of the Constellation-X Facility Science Team
(FST) for detailed discussions relating to the technical capabilities
of the mission, especially N.~White, H.~Tananbaum and R.~Mushotzky.
DR thanks the NASA Goddard Space Flight Center for hospitality during
the December 2006 Con-X FST meeting. We are grateful to A.~Jenkins for
sharing with us his code to calculate the mass function of dark matter
halos, and thank S.~Church and J.~Weller for discussions. We also
thank G.~Morris for technical support. The computational analysis was
carried out using the KIPAC XOC and Orange computer clusters at SLAC,
and the SLAC UNIX compute farm. SWA acknowledges support from the
National Aeronautics and Space Administration through Chandra Award
Number DD5-6031X issued by the Chandra X-ray Observatory Center, which
is operated by the Smithsonian Astrophysical Observatory for and on
behalf of the National Aeronautics and Space Administration under
contract NAS8-03060. This work was also supported in part by the U.S.
Department of Energy under contract number DE-AC02-76SF00515. AM was
additionally supported in part by a William~R. and Sara Hart Kimball
Stanford Graduate Fellowship.

\bibliography{biblist_simul}
\bibliographystyle{mn2e}

\label{lastpage}
\end{document}